\documentclass[12pt,letterpaper]{article}

\usepackage{graphicx,array}
\usepackage{url}
\usepackage{color}
\usepackage{latexsym}
\usepackage{amsthm}
\usepackage{amsmath}
\usepackage{amssymb}
\usepackage{amsfonts}
\usepackage[numbers,sort&compress]{natbib}
\usepackage{bm}
\usepackage{slashed}
\usepackage{mathrsfs}
\usepackage{enumerate}
\usepackage{tikz}
\usepackage{siunitx}
\usepackage{mdframed}
\usepackage{setspace}  
\usepackage{esvect}

\usepackage[utf8x]{inputenc}%
\usepackage{tcolorbox}%

\usepackage{hyperref} 
\hypersetup{
    colorlinks=true,       
    linkcolor=red,          
    citecolor=blue,        
    filecolor=magenta,      
    urlcolor=blue           
}
\usepackage[all]{hypcap} 

\usepackage{natbib}
\newcommand{\nc}{\newcommand}

\nc{\beq}{\begin{equation}}  
\nc{\eeq}{\end{equation}}  
\nc{\beqa}{\begin{eqnarray}}  
\nc{\eeqa}{\end{eqnarray}}  
\nc{\bit}{\begin{itemize}}  
\nc{\eit}{\end{itemize}}  

\def\TeV{\mathrm{TeV}}     
\def\GeV{\mathrm{GeV}}     
\def\cm{\mathrm{cm}} 

\newcommand{\eg}{{\it e.g.}}
\newcommand{\ie}{{\it i.e.}}
\newcommand{\Mpl}{M_{\rm pl}}

\newcommand{\cO}{\mathcal{O}}

\usepackage{floatrow}
\newfloatcommand{capbtabbox}{table}[][\FBwidth]

\usepackage{blindtext}

\newcommand\sbullet[1][1.5]{\textcolor[gray]{0.4}{ \mathbin{\vcenter{\hbox{\scalebox{#1}{$\ast\hspace{-1.8mm}{\raisebox{0.20ex}{\scriptsize$\bullet$}}$}}}} } }
\DeclareRobustCommand\encircle[1]{%
  \tikz[baseline=(X.base)]  \node (X) [draw, shape=circle, inner sep=-0.7] {\strut \raisebox{-0.5pt}[0pt]{#1}};}
\newcommand{\Mcirc}{\encircle{\textbf{M}}}

\newcommand{\Mmono}{M_{\tiny \Mcirc}}
\newcommand{\Rmono}{R_{\tiny \Mcirc}}
\newcommand{\Mewbh}{M_{\sbullet}}
\newcommand{\Newbh}{N_{\sbullet}}
\newcommand{\Mbh}{M_{\rm BH}}

\newcommand{\Qstopsun}{30}
\newcommand{\Mstopsun}{2 \times 10^{21}~\GeV}

\newcommand{\Qstopearth}{1900}
\newcommand{\Mstopearth}{1.2 \times 10^{23}~\GeV}

\newcommand{\BNV}{BNV}

\title{ 
	\vspace*{-2.3cm}  
	\begin{flushright}
		\normalsize{ \small PITT-PACC-2005
		}
	\end{flushright}
	\vspace{1.5cm}
{\bf Phenomenology of Magnetic Black Holes with Electroweak-Symmetric Coronas}
\author{\large Yang Bai$^{\,\star}$, Joshua Berger$\,^\diamond$, Mrunal Korwar$^{\,\star}$, and Nicholas Orlofsky$^{\,\star}$}
\date{\small \it 
$^\star$Department of Physics, University of Wisconsin-Madison, Madison, WI 53706, USA\\
$^\diamond$Department of Physics and Astronomy, University of Pittsburgh, Pittsburgh, PA 15260, USA \\
}
}

\begin{document}

\maketitle

\setlength{\parskip}{0.2ex}

\begin{abstract}	
Magnetically charged black holes (MBHs) are interesting solutions of the Standard Model and general relativity.  They may possess a ``hairy'' electroweak-symmetric corona outside the event horizon, which speeds up their Hawking radiation and leads them to become nearly extremal on short timescales.  Their masses could range from the Planck scale up to the Earth mass. We study various methods to search for primordially produced MBHs and estimate the upper limits on their abundance. We revisit the Parker bound on magnetic monopoles and show that it can be extended by several orders of magnitude using the large-scale coherent magnetic fields in Andromeda.  This sets a mass-independent constraint that MBHs have an abundance less than $6 \times 10^{-3}$ times that of dark matter.  MBHs can also be captured in astrophysical systems like the Sun, the Earth, or neutron stars.  There, they can become non-extremal either from merging with an oppositely charged MBH or absorbing nucleons. The resulting Hawking radiation can be detected as neutrinos, photons, or heat.  High-energy neutrino searches in particular can set a stronger bound than the Parker bound for some MBH masses, down to an abundance $10^{-7}$ of dark matter. 
\end{abstract}

\thispagestyle{empty}  
\newpage  
  
\setcounter{page}{1}  

\begingroup
\hypersetup{linkcolor=black,linktocpage}
\tableofcontents
\endgroup

\newpage

\section{Introduction}\label{sec:Introduction}
With the discovery of the Higgs boson in 
2012~\cite{Chatrchyan:2012ufa,Aad:2012tfa}, the complete particle content of 
the Standard Model (SM) of particle physics is confirmed. Although remaining 
puzzles such as neutrino mass, dark matter, and the baryon asymmetry may 
require physics beyond the SM, it is also important to know all possible states
of matter in the SM. For ordinary life, the electromagnetic interaction 
provides a rich ensemble of both stable and meta-stable atomic states. Quantum 
chromodynamics (QCD) interactions are responsible for hadronic states and 
atomic nuclei. In high-temperature and/or high-density environments, exotic 
states also exist based on the QCD interaction, such as quark matter or 
color-superconducting matter~\cite{Rajagopal:2000wf}, which could exist in the 
cores of neutron stars. For the SM electroweak (EW) interactions, less attention 
has been paid to the possible exotic states with the exception of objects 
similar to the `t Hooft-Polyakov 
monopole~\cite{tHooft:1974kcl,Polyakov:1974ek}. Due to the non-compactness of 
the $U(1)_Y$ symmetry, there is no localized, finite-energy monopole solution 
in the weak sector (see Refs.~\cite{Cho:1996qd,Bae:2002bm} for new interactions 
beyond the SM that facilitate a finite-energy solution and 
Ref.~\cite{Nambu:1977ag} for a metastable monopole-antimonopole state connected 
by a string). 

With the help of the gravitational interaction, more states based on the SM 
interactions could exist. One interesting example is the black hole with EW 
hair, discovered by Lee and Weinberg in Ref.~\cite{Lee:1994sk}. This 
magnetically charged black hole can have spherical non-topological monopole 
hair around a Reissner-Nordstr\"om (RN) black hole (BH) at the core. For a 
large magnetic charge $Q$, similar but much heavier objects exist. As recently 
pointed out by Maldacena in Ref.~\cite{Maldacena:2020skw}, EW symmetry can be unbroken inside a non-spherical corona-like region surrounding the BH. The restoration of EW symmetry in a large magnetic field 
was pointed out a while ago in 
Refs.~\cite{Salam:1974ny,Salam:1974xe,Ambjorn:1989bd,Ambjorn:1989sz,Ambjorn:1992ca}.
Other objects with macroscopic EW-symmetric (EWS) regions have been studied in \cite{Ponton:2019hux,Bai:2019ogh,Bai:2020ttp}.

This magnetic (nearly-)extremal black hole (MeBH) with an EWS corona serves as a novel and interesting state just from the SM plus general relativity. Near extremality, this new state is stable against Hawking radiation~\cite{Hawking:1974sw}. For a large enough $Q$, the  evaporation of magnetic charge via Schwinger production of magnetic monopoles like grand unified theory (GUT) monopoles or splitting into two smaller MeBHs is also suppressed. As a result, this MeBH with an EWS corona is stable on cosmological time scales, even at masses below $10^{15}\,\text{g}$ where uncharged BHs would evaporate on timescales shorter than the age of the Universe. Leaving aside the early universe production of primordial MeBHs (PMBHs) (see Refs.~\cite{Stojkovic:2004hz,Bai:2019zcd} for early universe production of primordial eBHs and Refs.~\cite{Sasaki:2018dmp,Carr:2020gox} for recent reviews of primordial uncharged BH production), the immediate questions are  how abundant MeBHs can be in the current Universe and how to detect them, which will be the main topics of this paper. 

From a phenomenological point of view, this MeBH has some similarities with the 
GUT monopole (see \cite{PDG2018,Mavromatos:2020gwk} for recent reviews). Both 
are heavy objects with magnetic charges, but MeBHs have a fixed, smaller charge-to-mass ratio and larger masses. Additionally, MeBHs could have 
a large-radius corona with unbroken EW symmetry, which will increase their 
interaction rates with ordinary matter. For instance, when we consider the 
capture rate of MeBHs by astrophysical objects, both the finite size and large magnetic charge of MeBHs will be taken into consideration. 
Furthermore, given the BH in the core, Hawking radiation also plays a role 
when either two MeBHs with opposite charges merge or one MeBH absorbs ordinary 
matter to become a non-extremal BH \cite{Bai:2019zcd}.  This Hawking radiation 
provides a unique signal not found in ordinary monopoles. As emphasized in 
Ref.~\cite{Maldacena:2020skw}, an interesting two-dimensional ($2d$) Hawking radiation mode could 
happen when the Hawking temperature is above the mass of the lightest 
electrically charged particle, \ie, the electron. PMBHs thus also differ from 
Schwarzschild BHs, which have four-dimensional ($4d$) Hawking radiation.

This paper is organized as follows. We first discuss some properties of MeBHs with an EWS corona in Section~\ref{sec:model}. In Section~\ref{sec:parker}, we apply the Parker limits to MeBHs from both the Milky Way and Andromeda galaxies.  We find that the large-scale coherent magnetic fields in Andromeda provide a much stronger bound on monopoles and MeBHs than the traditional Parker bound from the Milky Way. The neutrino signals from captured PMBHs inside the Sun are discussed in Section~\ref{sec:cosmic}; the Earth heat and neutrinos are worked out in Section~\ref{sec:earth}; the photon signals from the PMBHs captured by neutron stars and white dwarfs are studied in Section~\ref{sec:neutron-star}.  For some MeBH masses, several of these signals can provide even stronger constraints than the Andromeda Parker bound. We discuss other constraints like gravitational lensing and direct detection searches, then summarize the various constraints on the fraction of PMBHs as dark matter in Section~\ref{sec:conclusion}. In Appendix~\ref{sec:dirac-equations}, the formalism for $2d$ massless fermion modes in a PMBH magnetic field background and a qualitative understanding of the neutrino $2d$ modes are provided. In Appendix~\ref{sec:stopping}, stopping of a finite-sized PMBH in various astrophysical objects is worked out. We use natural units with $\hbar=c=\varepsilon_0=\mu_0=1$.

\section{Electroweak-symmetric corona black holes}\label{sec:model}

\subsection{Extremal magnetic black holes}

A charged black hole is described by the Reissner-Nordstr\"om metric, 
\beqa
\label{eq:metric-1}
ds^2 = f(r) \, dt^2 - \frac{dr^2}{f(r)} - r^2 \,d\theta^2 - r^2 \, \sin^2{\theta} \, d\phi^2 ~.
\eeqa
Here, $f(r) = 1 - 2 M G/r + \pi Q^2 G/(e^2 r^2)$; $G = 1/M^2_{\rm pl}$ with 
$M_{\rm pl} = 1.22 \times 10^{19}$~GeV; $e$ is the electric gauge coupling 
defined as $e=\sqrt{4\pi\alpha}$ with $\alpha\approx 1/137$ as the 
fine-structure constant; $Q$ is an integer and $h_Q \equiv Q\,h$ with $h = 
2\pi/e \approx 68.5\,e \approx 21$ is the magnetic charge times the coupling. 
$Q=1$ is the minimal charge, manifesting the Dirac quantization condition $e\,h = 2\pi$. The magnetic 
field is $\mathbf{B} = Q h\mathbf{\hat{r}}/(4\pi r^2)$. For a RN magnetic 
extremal black hole (eBH), one has 
\beqa
f(r) = \left( 1- \frac{R_\mathrm{eBH}^\mathrm{RN} }{r} \right)^2 \,, \qquad
M_\mathrm{eBH}^\mathrm{RN} = \frac{\sqrt{\pi} \, |Q|}{e} \, M_{\rm pl} \,, \qquad
R_\mathrm{eBH}^\mathrm{RN} = \frac{\sqrt{\pi}\,|Q|}{e}\, \frac{1}{M_{\rm pl}} ~.
\eeqa
Note that the repulsive magnetic force between two same-sign extremal BHs is exactly equal to the attractive gravitational force.  

Ref.~\cite{Lee:1994sk} demonstrated the existence of another 
magnetically-charged BH solution, which appears when the event horizon 
radius is less than of order the monopole radius.  By ``monopole radius,'' we 
mean the radius that an `t Hooft-Polyakov monopole would have if it was 
admissible in the symmetries of the SM, \ie ~$\Rmono \simeq m_W^{-1}$ when $Q=2$
with $m_W$ the $W$ boson mass.  This leads to a hairy BH with a cloud of EW 
fields outside the horizon, distinct from the RN solution.

The magnetic field near the event horizon of an eBH is 
\beqa
B(R_\mathrm{eBH}) = \frac{Q}{2\,e\,R^2_\mathrm{eBH}} \sim \frac{e\,M_{\rm pl}^2}{2\pi\,Q} ~,
\eeqa
where we have used $R_\mathrm{eBH} \sim R^\mathrm{RN}_\mathrm{eBH}$. For a 
smaller $Q$, the magnetic field is larger, owing to the shrinking event horizon 
radius. As studied in 
Refs.~\cite{Salam:1974ny,Salam:1974xe,Ambjorn:1989bd,Ambjorn:1989sz,Ambjorn:1992ca},
 EW symmetry is restored when the magnetic field satisfies $e B 
\gtrsim m_h^2$ where $m_h \approx 125$~GeV is the SM Higgs boson mass. The 
critical magnetic field is $B_{\rm EW} \approx m_h^2/e \approx 3\times 
10^{24}~\mbox{gauss}$, which provides an upper bound on $Q$ to have an 
EWS corona BH
\beqa
\label{eq:Qmax}
Q \lesssim Q_{\rm max}  \equiv \frac{e^2\,M_{\rm pl}^2}{2\pi\, m_h^2} \approx 1.4 \times 10^{32} ~.
\eeqa
Depending on physics beyond the SM, one could have a lower bound on $Q$ if 
there exist additional magnetic monopoles in the spectrum. For instance, if GUT monopoles exist, the EWS-corona BH may be 
Schwinger discharged by emitting GUT monopoles. The lower bound on $Q$ is 
(omitting a logarithmic factor, see 
\cite{Khriplovich:1999gm,Khriplovich:2002qn,Bai:2019zcd})
\beqa
\label{eq:Qmin-Schwinger}
Q \gtrsim  \frac{M_{\rm pl}^2}{\pi (\Mmono^{{\rm GUT}})^2 } \sim (5 \times 10^{3}) \left( \frac{10^{17}\,\mbox{GeV}}{\Mmono^{{\rm GUT}}} \right)^2 ~.
\eeqa

For a PMBH with $Q < Q_{\rm max}$, the EWS-corona radius within which $B>B_\text{EW} \approx m_h^2/e $ is roughly
\beqa
\label{eq:radius-ew}
R_\mathrm{EW} \simeq \sqrt{\frac{Q}{2}} \, \frac{1}{m_h}  ~.
\eeqa
Within this radius, the EW symmetry is unbroken. Analogously, there is also a QCD corona for the PMBH with radius
\beqa
R_\mathrm{QCD} \sim \sqrt{Q}\,\frac{1}{\Lambda_{\rm QCD} } ~,
\eeqa
where $\Lambda_{\rm QCD}\sim 1$~GeV. This radius is larger than 
$R_\mathrm{EW}$.  Within this radius, QCD may be in a different phase from the vacuum phase without a magnetic 
field~\cite{Shovkovy:2012zn,Kharzeev:2012ph}. 

Ref.~\cite{Lee:1994sk} demonstrated the existence of a spherically symmetric 
solution for $Q=2$ (corresponding to $q=1$ in \cite{Lee:1994sk}) for this new 
class of magnetically charged BHs with hair. The hairy BH's 
mass is at least the summation of $c_W M_\mathrm{eBH}^\mathrm{RN}$ and the 
monopole mass $\Mmono$. Here, $c_W=\cos{\theta_W} \approx 0.88$, $\theta_W$ is 
the Weinberg angle of the SM, and $\Mmono \simeq 4\pi\,m_W/e^2$ is the 
spherically symmetric monopole mass (again, assuming such a monopole was 
admissible in the SM symmetry group). The factor of $c_W$ appears because the 
EW symmetry is restored near the event horizon, so the BH carries magnetic
hypercharge $2\pi Q/g_Y=c_W 2 \pi Q/e$, with $g_Y$ the hypercharge 
coupling constant. Its mass is bounded from above by requiring the mass not be 
larger than that of a BH with radius $R_\text{EW}$. For a large $Q$, 
the corona boundary is anticipated to be non-spherical, and the mass $\Mewbh$ 
must be above $c_W M_\mathrm{eBH}^\mathrm{RN}$ plus the non-spherical 
$Q$-charged monopole mass $\Mmono(Q)$~\cite{Lee:1994sk}. The shape has not been worked out in detail, but may be expected to contain spiky features where vortex strings end on monopoles \cite{Maldacena:2020skw}, which we denoted using subscript$_{\sbullet}$.

We now give a more precise estimate for the mass. Including the contributions from both the hypercharged BH mass and the positive vacuum energy of the unbroken EW symmetry, $m_h^2\,v^2/8$, the EWS-corona BH mass is estimated to be
\beqa
M_\text{MeBH}^\text{tot}(Q)  &\simeq& c_W\, \frac{\sqrt{\pi}\,Q}{e}\,M_{\rm pl} \,+\, \frac{4\pi}{3}\,R_\mathrm{EW}^3\, \frac{m_h^2 \,v^2}{8} =  c_W\, \frac{\sqrt{\pi}\,Q}{e}\,M_{\rm pl} \,+ \, \frac{\pi}{12\sqrt{2}}\,Q^{3/2}\,\frac{v^2}{m_h} \\
&\equiv& \Mewbh(Q) +  \frac{\pi}{12\sqrt{2}}\,Q^{3/2}\,\frac{v^2}{m_h} ~,
\label{eq:mass-in-Q}
\eeqa
defining $ \Mewbh(Q) = c_W\, M_\text{eBH}^\text{RN}$. 
Here, we have ignored the energy contributions from the transition boundary from symmetry-unbroken to broken regions as well as the non-sphericity of the corona configuration. We anticipate that those corrections are small in the limit of $1 \ll Q \ll Q_{\rm max}$. The second term, which comes from the energy density of the corona, is only important when $Q \gtrsim 288 c_W^2/(\pi e^2) (\Mpl m_h/v^2)^2 \approx 5 \times 10^{35} \gg Q_\text{max}$, so we will generally neglect it.  

However, it is easy to see that $M_\text{MeBH}^\text{tot}(2\,Q) > 2\, M_\text{MeBH}^\text{tot}(Q)$ due to the 
presence of the second term, so energetically it is preferable for an MeBH with 
a large charge to split into smaller MeBHs. Although the large-charged MeBH is 
metastable, its lifetime can be longer than the age of the Universe for $Q 
\gtrsim Q_\text{min} \simeq 10^6$ given the existence of a GUT monopole with mass $\Mmono^{{\rm 
GUT}}\sim 10^{17}$~GeV~\cite{Maldacena:2020skw}. This is a stronger condition 
than in (\ref{eq:Qmin-Schwinger}). This metastability is in agreement with the 
weak gravity conjecture~\cite{ArkaniHamed:2006dz}: the non-gravitational 
interaction is stronger than the gravitational one.
The range of viable charges $Q_{\rm min} \lesssim Q \lesssim Q_{\rm max}$ 
corresponds to a mass range
\beqa
6 \times 10^{25}\,\mbox{GeV}\,\lesssim  &\Mewbh& \lesssim 9 \times 10^{51}\,\mbox{GeV}~.
\label{eq:mass-range} 
\eeqa
For reference, the mass of the Earth is $M_\oplus = 6.0 \times 10^{27}~\text{g} = 3.4 \times 10^{51}~\text{GeV}$.

\subsection{Non-extremal magnetic black holes}

Non-extremal BHs are also relevant for phenomenology. They appear, {\it e.g.}, 
after mergers of oppositely charged PMBH or absorption of baryons by PMBHs. For 
these cases, the BH mass $\Mbh > \Mewbh$, so the BH has a non-zero Hawking 
temperature given by
\begin{align}
\label{eq:temperature}
T(\Mbh, \Mewbh) = \frac{M^2_{\rm pl}}{2\pi}\, \frac{\sqrt{\Mbh^2 - \Mewbh^2} }{\left(\Mbh + \sqrt{\Mbh^2 - \Mewbh^2} \right)^2} ~.
\end{align}
Here and elsewhere, $\Mewbh$ is taken to mean only the mass contribution from the BH and not the corona as in (\ref{eq:mass-in-Q}).

For a non-extremal PMBH with an EWS corona, the Hawking radiation inside the 
corona is effectively made up of $2d$ modes (see Appendix 
\ref{sec:dirac-equations}) leading to a radiated 
power~\cite{1989JPhA...22.1073L,Maldacena:2020skw}
\beqa
\label{eq:2d-radiation}
P_2 = \frac{dE}{dt} = \frac{\pi\,g_*}{24}\,T^2(\Mbh, \Mewbh) ~.
\eeqa
Here, $g_*$ counts the number of left- and right-handed $2d$ modes using the 
hypercharges of chiral fermions. For instance, $g_* = |Q|$ for $q_L, 
\ell_L, d_R,  e_R$ (the left handed quark, lepton doublets, right handed down 
quark, and electron of the SM) and $g_* = 2 |Q|$ for $u_R$ (the right handed up 
quark). In the high-temperature limit, the total $g_* = 6 |Q|$ for one family 
of SM fermions and $g_* = 18 |Q|$ for three families. We emphasize that the 
$2d$ Hawking radiation only applies to fermions here (for spin-zero particles, 
the $2d$ modes are massive with a mass proportional to $\sqrt{q e 
B(R_\mathrm{eBH})}$, with $q$ the particle's charge; for spin-one particles, 
the magnetic flux generates a negative mass and induces gauge boson 
condensation), so no photon modes with a large multiplicity $|Q|$ are 
anticipated. Furthermore, not all of those fermion modes can travel outside of 
the EWS corona and be observed at a distant location. Electric-charged fermions 
can effectively travel to infinity if their energy is above their mass in the 
normal vacuum. Heavier particle emission with mass $m>T$ is suppressed by a 
Boltzmann factor of $e^{-m/T}$. For instance, when $m_e \lesssim T  \lesssim 
m_\mu$, only electrons can efficiently be $2d$ Hawking radiated and travel to 
infinity, and $g_*=2|Q|$ after taking into account both chiralities. 

Neutrinos do not have an electric charge and thus do not have $Q$-enhanced 
massless $2d$ modes outside the EWS corona. The $2d$ Hawking-radiated $\mathcal{O}(Q)$ 
neutrino modes around the event horizon are not able to freely travel outside 
the EWS corona (see Appendix~\ref{sec:dirac-equations} for more discussion). The 
characteristic energy barrier is $\mathcal{O}[\sqrt{e\,B(R_{\rm 
EW})}]=\mathcal{O}(m_h)$. For $T(\Mbh, \Mewbh)  \gtrsim m_h$ (which 
can be satisfied for $Q \lesssim M_{\rm pl}/m_h$ and $\Mbh$ not too close to $ 
\Mewbh$),
there are many other $Q$-enhanced modes for 
charged leptons and quarks, which can escape the EWS corona and directly (or 
after hadronization) decay into neutrinos. 

When $T(\Mbh, \Mewbh) \lesssim m_e$, the previous $2d$ radiation is suppressed. The region within the EWS corona will be heated to the Hawking temperature of the BH. Both thermal photon and neutrino modes are stored in this region. As a result, the $4d$ blackbody radiation on the boundary of the EWS corona could be important and has radiated power
\beqa
\label{eq:4d-radiation}
P_4 =  \frac{dE}{dt}  \approx \frac{\pi^2\,g_*}{120} (4\pi\,R_{\rm EW}^2)\,T^4(\Mbh, \Mewbh)~,
\eeqa
with $g_* = 2$ for photon and $g_* = 3\times 2 \times \frac{7}{8}=\frac{21}{4}$ for three chiral neutrinos. 
Eq.~(\ref{eq:4d-radiation}) is only valid for $T(\Mbh, \Mewbh) \lesssim m_e$. For a higher temperature, the radiated $2d$ modes can escape the EWS corona region without thermalizing with the center BH (even neutrinos can be converted to charged leptons plus gauge fields). 

When a pair of PMBHs merge, a non-extremal RN BH is generated.
The radius of a RN BH is $R_+ = (\Mbh + \sqrt{\Mbh^2 - \Mewbh^2})/M_{\rm pl}^2$. If two (near-)extremal PMBHs with charges $Q_1$ and $-Q_2$ satisfying $Q_1 \geq Q_2 > 0$ merge, the merger product has a charge of $Q = Q_1 - Q_2$ and mass of $\Mbh = \Mewbh(Q_1) +  \Mewbh(Q_2) \approx c_W \sqrt{\pi} (Q_1 + Q_2) M_{\rm pl}/e$. The condition, $e B(R_+)\gtrsim m_h^2$, for the merger product to have an EWS corona becomes $\sqrt{Q_1 - Q_2}/(\sqrt{Q_1} + \sqrt{Q_2})^2 > \sqrt{2\pi}\,c_W\,m_h/(e M_{\rm pl})=c_W/\sqrt{Q_\text{max}}$. So, unless $Q_1$ is infinitesimally close to $Q_2$ (or both are near $Q_\text{max}$), this condition can be easily satisfied and the PMBH merger product also has an EWS corona. In the limit of $Q_1 - Q_2 \ll Q_1 + Q_2 \equiv 2 Q$ and hence $M_\text{BH} \gg \Mewbh$, this condition becomes
\begin{equation}
\label{eq:Q1Q2-corona-cond}
\Mbh < \frac{M_{\rm pl}^2}{2\sqrt{2}\,m_h}\,\sqrt{Q} = \frac{\sqrt{\pi}}{2\,e}\,\sqrt{Q\,Q_{\rm max}}\,M_{\rm pl} \equiv M^{\rm EW}_{\rm max}(Q) ~. 
\end{equation}
Thus, the produced non-extremal RN BH also has an EWS corona (we do not consider the situation that the charge distribution of PMBHs is exactly monochromatic with a delta function). Using \eqref{eq:temperature}, the Hawking temperature is
\beqa
\label{eq:TBH-merger}
T_{\rm BH} \simeq \frac{M_{\rm pl}^2}{2\pi}\, \frac{1}{8\,\Mewbh(Q_1)} = (2.8 \times 10^{10}\,\mbox{GeV})\, M_{26}^{-1} ~,
\eeqa
where $M_{26}=\Mewbh/10^{26}\,\mbox{GeV}$. 
For $T_\text{BH}>m_e$, \ie~when $\Mewbh \lesssim 10^{39}\,\mbox{GeV}$, the $2d$ 
radiation in \eqref{eq:2d-radiation} dominates. For a smaller mass $\Mewbh 
\lesssim 10^{37}\,\mbox{GeV}$, muons and other electrically charged particles 
can be produced from the $2d$ Hawking radiation. The heavy charged particles 
have various decay channels which generate neutrinos that can escape the Sun or 
the Earth's core and potentially be observed by neutrino telescopes. In 
addition, the radiated particles other than the neutrinos may heat up 
astrophysical bodies like the Earth, neutron stars, or white dwarfs, 
potentially providing a bound. Also note that the non-extremal RN BH can 
quickly $2d$ Hawking radiate to become (nearly) extremal. Using 
\eqref{eq:2d-radiation}, the evaporation time scale is 
\beqa
\label{eq:tau-merger}
\tau_{\rm BH} \approx \frac{3000\,\pi^{3/2}\,c_W}{e}\,\frac{\Mewbh^2}{M_{\rm pl}^3} \approx (1.8\times 10^{-25}\,\mbox{s})\, M_{26}^2 ~,
\eeqa
where we have chosen $g_* = 2 |Q|$ and $\Mbh =2 \Mewbh$. As emphasized in Ref.~\cite{Maldacena:2020skw}, this $2d$ Hawking radiation time scale is much shorter than the $4d$ one, which scales like $\Mewbh^3/M_{\rm pl}^4$~\cite{Bai:2019zcd}.

PMBHs can also facilitate baryon number violation (\BNV).  As discussed in Appendix \ref{sec:dirac-equations}, baryons that enter an EWS corona become $2d$ modes and can be easily captured by the PMBH.  For example, if $R_\text{EW} \gtrsim 1~\GeV^{-1}$ corresponding to $Q \gtrsim 10^4$, then baryon bound states are expected \cite{Bai:2019ogh}.  The PMBH can then reemit energy as Hawking radiation, which need not have the same baryon number.
There is also the possibility that the EWS corona can mediate baryon 
number violation \cite{Arnold:1987mh,Ho:2020ltr}. We do not consider that 
here because the extended sphaleron configuration is also relevant for the 
baryon-violating process, though the sphaleron energy is 
reduced~\cite{Ho:2020ltr}.
Note that these processes are different from the Callan-Rubakov process, which applies to GUT monopoles 
\cite{Rubakov:1981rg,Rubakov:1982fp,Callan:1982ah,Callan:1982au,Callan:1982ac}.
Another interesting possibility, left for future work, is that the \BNV~process could facilitate baryogenesis.

For the case of PMBH absorption of baryons, the resulting BH mass is close to the extremal mass. In the limit of $\Mbh - \Mewbh \ll \Mewbh$, the Hawking temperature is
\beqa
\label{eq:TBH-smallDM}
T_{\rm BH} \simeq \frac{M_{\rm pl}^2}{\sqrt{2}\,\pi}\,\frac{\sqrt{\Mbh - \Mewbh}}{\Mewbh^{3/2} } ~.
\eeqa
For the $2d$ evaporation process to occur, $T_{\rm BH} \gtrsim m_e$ or $\Mbh-\Mewbh \gtrsim 2\pi^2m_e^2\Mewbh^3/M_{\rm pl}^4 \approx (2.5 \times 10^{-4}) m_p M_{26}^{3}$.  For example, when even a single proton is absorbed ($\Mbh-\Mewbh \simeq m_p$), the $2d$ evaporation process occurs for $\Mewbh \lesssim 10^{27}$~GeV, resulting in a prompt \BNV~process. PMBHs with larger masses must absorb many baryons before reemitting via $2d$ modes.  This may occur, \eg, in dense environments like stars. Using the $2d$ radiation in \eqref{eq:2d-radiation} with $g_*=2|Q|$, the fast $2d$ evaporation time scale is 
\beqa
\tau_{\rm BH} \approx \frac{24\pi^{3/2}\,c_W\,\Mewbh^2}{e\,M_{\rm pl}^3}\, \log{\left[ \frac{M_{\rm pl}^4\,(\Mbh-\Mewbh)}{2\pi^2\,m_e^2\,\Mewbh^3}  \right]} ~.
\eeqa
After this time scale, the BH follows the slow $4d$ evaporation process.

\section{Parker limits from Milky Way and Andromeda galaxies}\label{sec:parker}

The Parker bound arises from the requirement that domains of coherent magnetic 
field are not drained by magnetic monopoles~\cite{Parker:1970xv}.  If a 
monopole transits such a domain, it will be accelerated by the magnetic field 
and drain its energy.  Thus, the energy loss to monopoles must be slower than 
the time it takes for the fields to be regenerated. To simplify our discussion, 
we will ignore the subleading second term in \eqref{eq:mass-in-Q} for $Q \ll 
Q_{\rm max}$ and take $\Mewbh/Q = c_W\sqrt{\pi}M_{\rm pl}/e \approx 5.1\,M_{\rm 
pl}$. Compared to a GUT monopole with $Q=1$, a PMBH has a much larger 
mass-to-charge ratio. We now compare the PMBH flux to the various Parker-type 
bounds, updated to include charge dependence where necessary.

Assuming that PMBHs account for a faction $f_{\sbullet}$ of all dark matter energy density and has an averaged speed $v$, the flux is 
\beqa
\label{eq:local-flux}
F_{\sbullet} \approx (9.5 \times 10^{-21}\,\mbox{cm}^{-2}\mbox{sr}^{-1}\mbox{s}^{-1})\,f_{\sbullet}\, \left( \frac{10^{26}\,\mbox{GeV}}{\Mewbh} \right)\left( \frac{\rho_{\rm DM}}{0.4~\mbox{GeV}\,\mbox{cm}^{-3}}  \right) \left( \frac{v}{10^{-3}} \right) ~.
\eeqa
For the local dark matter density in our solar system, we use $\rho_{\rm local} \approx  0.4~\mbox{GeV}\,\mbox{cm}^{-3}$~\cite{Karukes:2019jxv} and virial velocity $v \approx 10^{-3}$~\cite{Griest:1986yu}.

We follow the treatment of Ref.~\cite{Turner:1982ag}, but include the 
$Q$-dependence in $h_Q$ and $\Mewbh$. First, a monopole can be accelerated in a 
coherent magnetic field to reach a speed
\begin{align}
\label{eq:Parker-vmag-1}
v_{\rm mag} &\simeq \min \left[1, ~\sqrt{\frac{2\,B\,h_Q\,\ell_c}{\Mewbh}} \,\right] \simeq 4 \times 10^{-5}\, \sqrt{\ell_{21} B_3} ~,
\end{align}
where $\ell_{21}=\ell_c/(10^{21}~\cm)$ is the coherence length of the magnetic field and $B_3=B/(3\times 10^{-6}~\text{gauss})$ is the magnetic field strength in our Milky Way galaxy~\cite{Han2018}. This velocity is less than the virial velocity of our galaxy, around $10^{-3}$. Thus, the PMBHs can remain bound in our galaxy and explain DM.  They could also have a larger velocity and not be bound, thus unable to explain DM, but the flux bound presented below turns out to be the same. 

The Parker bound is set by requiring the mean rate of energy gained by PMBHs times the regeneration time $t_\text{reg}$ of the field by dynamo action  to be smaller than the energy stored in the magnetic field, or \footnote{The energy density of the magnetic field is $B^2/(2\mu_0)=B^2/2$ in natural units used here, differing from the units in \cite{Turner:1982ag}.}
\beqa
\Delta E \times F_{\sbullet} \times (\pi \ell_c^2)\times (4\pi~\text{sr})\times t_{\rm reg} \lesssim \frac{B^2}{2}\, \frac{4\pi\,\ell_c^3}{3}~,
\eeqa
with $\Delta E \simeq \Mewbh \,\Delta v^2 /2$ and $\Delta v \simeq B\,h_Q\,\ell_c/(\Mewbh v)$. The magnetic-field-independent constraint on the PMBH flux is 
\beqa
\label{eq:FParker-1}
F_{\sbullet} \lesssim (4.8\times 10^{-19}\,\mbox{cm}^{-2}\mbox{sr}^{-1}\mbox{s}^{-1}) \, \frac{v_{-3}^2}{\ell_{21}\,t_{15}\,M_{26}} ~,
\eeqa
where $v_{-3} = v/(10^{-3})$ and $t_{15}=t_{\rm reg}/(10^{15}\,\mbox{s})$. Combined with \eqref{eq:local-flux}, the constraint on the PMBH fraction from coherent fields in the Milky Way is independent of the PMBH mass and given by
\beqa
\label{eq:Parker-density-bound}
f_{\sbullet} \lesssim 50 \times \frac{v_{-3}}{\rho_{0.4}\,\ell_{21}\,t_{15} } ~,
\eeqa
where $\rho_{0.4} = \rho_{\rm DM}/(0.4\,\mbox{GeV}\,\mbox{cm}^{-3})$.

Thus, at present there is no constraint from magnetic field domains in our galaxy, regardless of PMBH mass and charge. To strengthen the bound in (\ref{eq:Parker-density-bound}),  one could look for systems with larger coherent magnetic field domains $\ell_{21} > 1$, longer times to regenerate the magnetic fields $t_{15}>1$, smaller virial velocities $v_{-3}<1$ (although note if $v<v_\text{mag}$, then PMBHs would not be bound to the galaxy so could not be DM, and a different constraint would apply \cite{Turner:1982ag}), or larger enhancements to the local DM density $\rho_{0.4}>1$. 

We identify the Andromeda galaxy as an example of a system with larger coherent magnetic domains that take a correspondingly longer time to regenerate. Andromeda has an approximately azimuthal magnetic field around its whole circumference, measured between radii of 6 and 14 kpc \cite{Fletcher:2003ec}. \footnote{The coherent magnetic field geometry for Andromeda is cylindrical, slightly different from the spherical geometry of domains in the Milky Way. We neglect this $\mathcal{O}(1)$ factor.}  This implies $\ell_c \sim 10~\text{kpc} \Rightarrow \ell_{21} \sim 30$ and $t_{\rm reg} \sim 10~\text{Gyr}  \Rightarrow t_{15} \sim 300$ \cite{Arshakian:2008cx}.  The density of DM for Andromeda is very similar to the Milky Way \cite{Klypin:2001xu,Tamm:2012hw}, so we keep $\rho_{0.4} \approx 1$ and $v_{-3} \approx 1$.
Using these values in (\ref{eq:Parker-density-bound}), we constrain the PMBH fraction in Andromeda to be 
\beqa
\label{eq:parker-limit-M31}
 \qquad \qquad \qquad f_{\sbullet} \lesssim 6 \times 10^{-3} \, \qquad \qquad  \mbox{(from M31)} ~. 
\eeqa
Although there is a large uncertainty for $\ell_c$ and $t_{\rm reg}$ used in the Parker limit, the above limit suggests PMBHs cannot account for all dark matter in our Universe. 

While the above bounds come from galactic magnetic fields, intracluster magnetic fields were considered in Ref.~\cite{Rephaeli:1982nv}, although the bound is somewhat less secure as stated in their paper. Because of the smaller intracluster dark matter density  $\approx 1.5\times 10^{-6}~\mbox{GeV}\,\mbox{cm}^{-3}$~\cite{Aghanim:2018eyx}, the constraint is much weaker than \eqref{eq:parker-limit-M31}. Finally, the bound in \cite{Turner:1982ag} was extended in \cite{Adams:1993fj}, which observed that the much smaller seed magnetic fields early in our galaxy's formation must also survive. Otherwise, there would be no fields today. The limit from the progenitor field reduces to the ordinary Parker bound when $\Mewbh/Q > 10^{17}~\GeV$, which is always satisfied for PMBHs.

\section{Cosmic rays: solar neutrinos}\label{sec:cosmic} 

PMBHs can be captured by the Sun, drift into the core of the Sun, and merge to produce non-extremal RN BHs that Hawking radiate energetic neutrinos. This high energy particle signal from merging extremal BHs was pointed out in a different context in \cite{Bai:2019zcd}.  In addition, PMBHs that are captured but have not yet annihilated can mediate \BNV~processes, which could also be detected by the energetic particles they emit. Measuring the neutrino flux from the Sun's direction by a large-volume neutrino detector can therefore constrain the PMBH fraction $f_{\sbullet}$ of dark matter. 

Before giving the detailed calculations, we provide a brief overview of the capture process, which will be applicable for the Sun, the Earth, neutron stars, and white dwarfs.  First, the Sun captures PMBHs with a rate dependent on the flux and the strength of interactions between PMBHs and the stellar medium.  The PMBHs will drift to the core of the Sun, and the time it takes should be short compared to other timescales in the problem.  There, they may merge with oppositely-charged PMBHs and ``annihilate'' via Hawking radiation.  The capture and annihilation rates often reach an equilibrium, so that the annihilation rate saturates to the capture rate.  
However, magnetic fields in the core may separate oppositely-charged PMBHs, preventing them from annihilating.  These can lead to a build-up of PMBHs, and these PMBHs can mediate \BNV~processes.

\subsection{PMBH capture by the Sun}\label{sec:solar-capture}

The capture rate of PMBHs by the Sun is estimated to be 
\beqa
\label{eq:capture-rate}
C_{\rm cap} \approx  \epsilon\,\pi \, R_\odot^2 \left[ 1 + (v_{\rm esc}/v)^2\right] \,  4\,\pi F_{\sbullet} \approx \left(9.2\times 10^{3}\,\mbox{s}^{-1} \right) \, \epsilon\,f_{\sbullet}\, M_{26}^{-1}\, ,
\eeqa
where $R_\odot = 7.0 \times 10^{10}\,\mbox{cm}$ is the solar radius; $v_{\rm 
esc} = 2\times 10^{-3}$ is the solar escape speed on the solar surface; $v = 
10^{-3}$ is the averaged dark matter speed; $\epsilon \in [0,1]$ is the capture 
efficiency parameter.  For $F_{\sbullet}$ in \eqref{eq:local-flux} we use 
$\rho_{\rm DM} = \rho_{\rm local} \approx 0.4~\mbox{GeV}\,\mbox{cm}^{-3}$. For 
a small radius PMBH, its stopping power by the solar plasma is similar to the 
GUT monopole 
case~\cite{Hamilton:1984rh,Meyer:energyloss,Frieman:1985dv,Ficenec:1987vy,Ahlen:1996ax}
 except for a factor of $Q^2$ enhancement. Using the results 
of~\cite{Ahlen:1996ax}, any PMBH with $Q> \Qstopsun$ or $\Mewbh \gtrsim 
\Mstopsun$ is stopped. Following the analysis of Ref.~\cite{Meyer:energyloss}, 
in Appendix~\ref{sec:stopping}, we also take into account finite-size effects 
and demonstrate that our Sun can easily stop any large-radius PMBH  above this 
minimal charge once it enters the Sun. We therefore choose $\epsilon=1$. 

The captured PMBH will be slowed down by interacting with the plasma, 
thermalize with the medium, and drift into the core region of the Sun. 
Oppositely charged PMBHs ``annihilate'' or merge into a non-extremal RN BH. The 
annihilation or merger rate is related to the number density distribution of 
PMBHs. 

We begin by presenting the usual calculation of solar capture and annihilation, relevant for DM with no self interactions aside from annihilations~\cite{Jungman:1995df}.  After that, we will include the effects of the magnetic fields of the Sun and PMBHs, which qualitatively and quantitatively change the results.  For the non-interacting case, the PMBH radial distribution at the core follows a Maxwell-Boltzmann distribution $\propto e^{-\phi(r)/(k_{\rm B} T)}$ where $\phi(r)$ is the gravitational potential. This can be written as a Gaussian with a characteristic radius of $R$:
\beqa
n_{\sbullet}(r) \, =\, n_0 \, e^{- r^2 /R^2} ~, 
\eeqa
with $n_0$ as the PMBH number density at the center of the Sun. The annihilation or merger rate is estimated to be
\beqa
C_A \approx  \frac{\int d^3r\,n(r)^2\,\langle \sigma_A \, v\rangle }{[\int d^3r\, n(r)]^2} \, \simeq  \, \frac{\pi\,R_{\rm eBH}^2}{(2\pi)^{3/2}\,R^3} ~,
\eeqa
where a geometric cross section with the BH event horizon radius is used. The 
radius $R$ can be estimated using the balance of the gravitational potential 
and kinetic energy $\frac{3}{2}k_{\rm B}\,T = \phi(r) = \frac{2\pi}{3}\, G 
\rho_c\,\Mewbh\,r^2$, where $\rho_c$ is the mass density in the core. If $T$ is 
similar to the solar core temperature without PMBHs, $T=T_c=1.5\times 
10^7$~K. If the gravitational potential is dominated by the solar plasma and 
using $\rho_c = \rho_p  \approx 
50\,\mbox{g}/\mbox{cm}^3$~\cite{solar-core-density}, the radius 
is~\cite{Griest:1986yu} 
\beqa
\label{eq:Rth}
R \approx R_{\rm th} \approx 0.13\,R_\odot\, \sqrt{ \frac{m_p}{\Mewbh} } \, =\, (8.8 \times 10^{-4}\,\mbox{cm})\, M_{26}^{-1/2} ~,
\eeqa
with the proton mass $m_p = 0.938$~GeV. The annihilation rate is given by
\beqa
\label{eq:CA-sun}
C_A  \simeq\frac{\pi\,R_{\rm eBH}^2}{(2\pi)^{3/2}\,R_{\rm th}^3} \approx (1.7\times 10^{-33}\,\mbox{s}^{-1})\, M_{26}^{7/2} ~. 
\eeqa

The time evolution of the PMBHs in the Sun is given by $\dot{N}_{\sbullet} = C_{\rm cap} - C_A \, \Newbh^2$~\cite{Jungman:1995df}, where all PMBHs are assumed to have charges of equal magnitude. The solution to this number evolution equation is $\Newbh(t) = \sqrt{C_{\rm cap}/C_A  }\, \tanh\left({\sqrt{C_{\rm cap}\, C_A}\,t}\right)$, which has an equilibration time
\beqa
\label{eq:tau-eq}
\tau_{\rm eq} = 1/\sqrt{C_{\rm cap}\, C_A} = (2.5\times 10^{14}\,\mbox{s})\, f_{\sbullet}^{-1/2}\, M_{26}^{-5/4}~. 
\eeqa
So, for $f_{\sbullet}=1$ and $\Mewbh > 6\times 10^{23}$~GeV, $\tau_{\rm eq}$ is shorter than the age of the Sun $t_\odot = 4.6\times 10^9\,\mbox{yr} = 1.45 \times 10^{17}\,\mbox{s}$. For $\tau_{\rm eq} < t_\odot$, the annihilation rate is determined by the capture rate and given by
\beqa
\label{eq:annihilation-rate}
\Gamma_A = \frac{1}{2}\,C_A\,\Newbh^2  \approx \frac{1}{2}\, C_{\rm cap}\,  = \, \left(4.6\times 10^{3}\,\mbox{s}^{-1} \right) \,f_{\sbullet}\, M_{26}^{-1}~. 
\eeqa
The time between mergers $\Gamma_A^{-1}$ is longer than $\tau_\text{BH}$ in 
(\ref{eq:tau-merger}) for $\Mewbh \lesssim 10^{39}\,\mbox{GeV}$, so one could 
approximately treat the $2d$ Hawking radiation as occurring instantaneously 
below this mass.  In other words, for low enough PMBH mass, PMBH mergers and 
annihilations can be thought of interchangeably.  We will see that all masses 
for which constraints can be placed satisfy this approximation. The total 
number of PMBHs inside the Sun is then
\beqa
N_{\sbullet} \approx \sqrt{\frac{C_{\rm cap}}{C_A}} \simeq (2.3 \times 10^{18})\,\left(\frac{R}{R_{\rm th}}\right)^{3/2}\,f_{\sbullet}^{1/2}\,M_{26}^{-9/4} \, .
\eeqa

So far, the PMBHs were assumed to make their way to the center of the Sun 
nearly instantaneously.  However, if the drag forces are large enough, it may 
take a long time for PMBHs to make their way from the surface to the center of 
the Sun.  An estimate for this drift velocity can be obtained by setting the 
stopping force in Appendix \ref{sec:stopping} equal to the gravitational 
attraction:
\beqa
\frac{G \Mewbh M_\odot(r)}{r^2} \sim v_\text{drift} \frac{n_e\,e^2\,h_Q^2}{4\pi\,m_e\,v_\text{th}} \, .
\eeqa
Thus, using $n_e=10^{24}~\cm^{-3}$ and $v_\text{th}=0.058$ corresponding to $T=10^7~\text{K}$, the drift time (for a vertical moving PMBHs) is 
\beqa
\label{eq:tdrift}
t_\text{drift} \sim \frac{R_\odot}{v_\text{drift}} &\sim & \frac{R_\odot^3}{M_\odot} \frac{n_e\, e^2}{c_W^2\,m_e\,v_\text{th} } \Mewbh  \sim (8\times 10^{4}\,\mbox{s})\,M_{26} ~. 
\eeqa
Thus, PMBHs with mass $\Mewbh \lesssim 2 \times 10^{38}~\GeV$ will take shorter than $t_\odot$ to settle to the center of the Sun.  This sets an upper limit on the masses that can be probed by PMBH merger signals.

The above analysis has ignored the magnetic field in the core of Sun. A magnetic field could separate the locations of positively and negatively charged PMBHs, as emphasized in Ref.~\cite{Frieman:1985dv}.  This would prevent oppositely charged PMBHs from merging unless the number of captured PMBHs is above a critical value.  On the other hand, the attractive forces between oppositely charged monopoles can increase the annihilation rate well above that in (\ref{eq:CA-sun}).  As we now demonstrate, the attractive forces are more important for the Sun when $f_{\sbullet}$ is not too small, while for tiny $f_{\sbullet}$ the Sun's magnetic field gives the dominant effect.  In either case, $\tau_\text{eq}$ in (\ref{eq:tau-eq}) will not be meaningful.

\begin{figure}[t]
	\centering
	\includegraphics{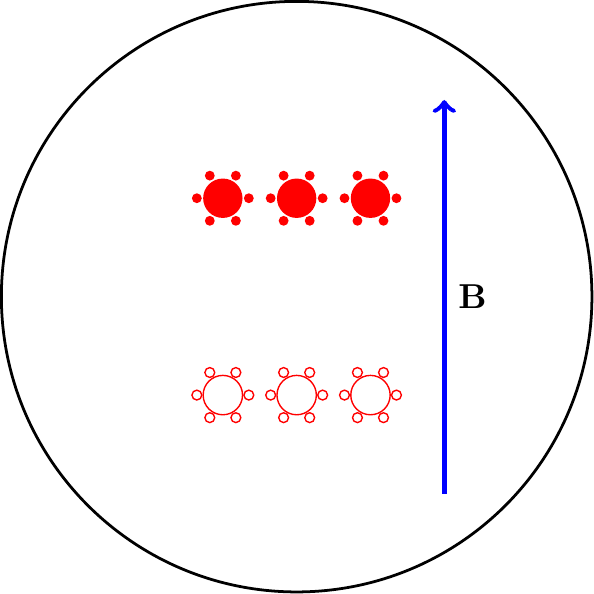}
	\caption{Separation of PMBHs (filled red) and anti-PMBHs (hollow red) in the Sun generated by a roughly uniform magnetic field $\mathbf{B}$ (blue).}\label{fig:sun-separation}
\end{figure}

Choosing a constant $B$ and assuming the $+Q$ and $-Q$ PMBHs are separated by a distance $z$ from the center, the force balance equation is
\beqa
\label{eq:force-balance}
0= F = B\,\frac{2\pi Q}{e} \, -\, \frac{4\pi}{3}\,G\,\rho_c\,\Mewbh\,z - \frac{G\,N_{\sbullet}\,\Mewbh^2}{(2\,z)^2}  ~,
\eeqa
where the last term is the attractive force of $\frac{1}{2}N_{\sbullet}$ PMBHs on one oppositely charged PMBH. We have assumed that the PMBHs are near-extremal so that the gravitational and magnetic forces are equal (they actually differ slightly, by a factor of $c_W$, which we neglect). If the first two terms are dominant, the magnetic and gravitational forces can be balanced with a separation distance
\beqa
z_B \simeq \frac{3\,B\,M_{\rm pl}}{2\sqrt{\pi}\,c_W\,\rho_c} = (2.0\times 10^3\,\mbox{cm})\, B_{100}~,
\eeqa
where $B_{100} = B/(100\,\mbox{gauss})$ (the precise value of magnetic field in 
the solar core region is still unknown~\cite{Solanki:2010je}) and $\rho_c = 
\rho_p  \approx 50\,\mbox{g}/\mbox{cm}^3$. Compared to $R_{\rm th}$ in 
\eqref{eq:Rth}, the magnetic force is more important than the thermal pressure 
and can potentially separate oppositely charged PMBHs and reduce the 
annihilation rate $C_A$ (as shown in Fig.~\ref{fig:sun-separation}). To 
consistently ignore the attractive force between PMBHs in the third term, 
the total number of PMBHs is required to be below a critical value
\beqa
\label{eq:N-critical}
N^{\rm crit}_{\sbullet} \simeq \frac{18\,M_{\rm pl}^3\,B^3}{\sqrt{\pi}\,c_W^3\,\Mewbh\,\rho_c^2} 
= (3.8 \times 10^{10})\,B_{100}^3\, M_{26}^{-1} ~.
\eeqa
Above this critical number, the third term in \eqref{eq:force-balance} reduces the distribution radius, and equilibrium is quickly reached between capture and annihilation with $C_{\rm cap} = C_A \,(N^{\rm crit}_{\sbullet})^2$. Starting from time zero, the captured PMBHs settle down in the solar core with a separation distance of $z_B$. The amount of time to reach $N^{\rm crit}_{\sbullet}$ is given by
\beqa
\label{eq:tau-crit}
\tau^{\rm crit} = \frac{N^{\rm crit}_{\sbullet}}{C_{\rm cap} } \simeq (4.1\times 10^6\,\mbox{s})\,B_{100}^3\,f_{\sbullet}^{-1} ~,
\eeqa
which is independent of PMBH mass and dramatically smaller than the age of the Sun unless $f_{\sbullet} \lesssim 3 \times 10^{-11}$.

Note that when $10^{36}~\mbox{GeV} \lesssim \Mewbh \lesssim 2\times 10^{38}~\mbox{GeV}$, there are only $\mathcal{O}(1)$ PMBHs within the distribution radius to annihilate. Also, for a heavy mass, the capture rate in \eqref{eq:capture-rate} is small. The capture/annihilation process is discrete: the Sun waits a long time to capture the next PMBH, which quickly drifts into the core and annihilates with the existing one.

\subsection{Solar neutrinos from PMBH annihilation}\label{sec:solar-neutrino}

For the radiated charged fermions following a PMBH merger, the thermally 
averaged energy for $2d$ Hawking radiation is $\langle E_f \rangle \approx 1.19\,T_{\rm BH}$, with 
$T_\text{BH}$ given in (\ref{eq:TBH-merger}). Ignoring order one factors, the 
number of charged particles from one PMBH annihilation event is $N_{f} \approx 
\Mewbh/T_{\rm BH}$. Charged particles like $\mu^\pm$, $\pi^\pm$, and $K^\pm$ can 
decay into neutrinos directly or via cascade. The total number of high energy 
neutrinos from each annihilation event is 
\beqa
N_\nu \approx \eta_\nu\,  \frac{\Mewbh}{T_{\rm BH}} =  (3.4\times 10^{15})\,\eta_\nu\,M_{26}^2 ~,
\eeqa
where $\eta_\nu$ is a factor to take into account the average number of high 
energy neutrinos from charged particle decays, {\it e.g.}, $\eta_\nu \approx 2$ 
for a muon.  This assumes that the particles do not thermalize or significantly 
slow down in the solar plasma before decaying to neutrinos, which is true for 
prompt-decaying particles like the $\tau$ lepton, as well as the charm and 
bottom hadrons.
The neutrino energy is around $E_\nu \simeq \langle E_f \rangle / \eta_\nu \approx (1.19/\eta_\nu) \,T_{\rm BH}$ assuming $\eta_\nu \geq 1$. The neutrinos generated from PMBH annihilation plus $2d$ Hawking radiation can propagate outside the Sun and reach the Earth to be observed by neutrino telescopes. 

IceCube has performed a search for dark matter annihilations inside the 
Sun~\cite{Aartsen:2016zhm}, which can be recast as a search for PMBHs. There 
are lower energy cuts to select neutrinos: $E_\nu^\text{cut}= 10$ GeV for 
DeepCore selection and 100~GeV for IceCube selection. Requiring $T_{\rm BH} > 
E_\nu^\text{cut}$ in \eqref{eq:TBH-merger}, the maximum mass that can be probed is
\beqa
\label{eq:upper-mass-energy}
\Mewbh \lesssim M_{{\rm max}, E}  = (2.8\times 10^{35}~\mbox{GeV}) \left( \frac{10\,\mbox{GeV}}{E_\nu^{\rm cut}} \right) ~. 
\eeqa
For a  heavy PMBH, the annihilation rate could be so suppressed such that the separation time from one event to another event could be longer than the operation time, $t_{\rm exp}$, of the experiment.  This sets another upper limit on the mass that can be probed
\beqa
\label{eq:upper-mass-time}
\Mewbh \lesssim M_{{\rm max}, t}  = (2.1\times 10^{37}~\mbox{GeV})\,f_{\sbullet}\, \left( \frac{t_{\rm exp}}{532\,\mbox{day}} \right) ~. 
\eeqa
For a given experiment, the combined upper limit on the testable mass is 
\beqa
\Mewbh \lesssim \mbox{min}\left[ M_{{\rm max}, E},  M_{{\rm max}, t}  \right] ~.
\eeqa

The generated neutrino flux is
\beqa
I_{\rm \nu} \approx \frac{N_\nu\, \Gamma_A}{4\pi\,d_\oplus^2} \approx (5.5\times 10^{-9}\,\mbox{cm}^{-2}\,\mbox{s}^{-1})\,M_{26}\,\eta_\nu\,f_{\sbullet} ~,
\eeqa
where the distance from the Earth to the Sun is $d_\oplus=1.5\times 
10^{13}\,\mbox{cm}$. The Sun is opaque for neutrinos with $E_\nu \gtrsim 
100$~GeV~\cite{Gandhi:1995tf}, so the high energy neutrinos generated at the 
solar core will be converted to charged leptons from the charged-current 
interaction or to lower-energy neutrinos from the neutral-current interaction. 
A detailed analysis requires a numerical study of particle production, decay, 
and interaction. In our simplified recast of IceCube limits, we take the 
initial neutrino energy $E_\nu \sim T_\text{BH}$. When $T_\text{BH}$ below 100 
GeV, we will take $\eta_\nu = 1$. For $T_\text{BH}\gtrsim 100$~GeV, we assume 
that the neutrinos with energy below or around 100 GeV have a multiplicity 
factor proportional the neutrino energy, $\eta_\nu \sim (T_{\rm 
BH}/100\,\mbox{GeV})$, resulting from the cascades of higher energy particles. 

For the IceCube searches~\cite{Aartsen:2016zhm}, $t_{\rm exp} = 532$~day. 
The effective area can be read from the left panel of Fig.~4 of  Ref.~\cite{Aartsen:2016zhm}, which can be approximated by
\begin{equation}
\renewcommand{\arraystretch}{1.3}
A_{\rm eff} \approx 
\left(1.0\times 10^{-2}\,\mbox{cm}^2\right) \times (E_\nu / 10\,\mbox{GeV})^3 \qquad  \quad 10 < E_\nu/\mbox{GeV} \leq 100  \,,
\end{equation}
for the acceptance area for the DeepCore selection. We also note that the search in Ref.~\cite{Aartsen:2016zhm} is for muon neutrinos, which generate tracks in the detector with a better pointing ability. Here, we absorb the additional neutrino flavor dependence and also neutrino oscillation effects into the factor $\eta_\nu$. Requiring the number of  signal events $I_\nu \times A_{\rm eff} \times t_{\rm exp} \lesssim 100$ (see Fig.~6 of \cite{Aartsen:2016zhm} for the observed number of events and Table 3 for systematical errors), we derive an approximate upper limit on the fraction of dark matter as PMBH
\beqa
\label{eq:bound-sun-IC}
\renewcommand{\arraystretch}{1.5}
f_{\sbullet} \lesssim \left\{
\begin{array}{ll}
1.4 \times 10^{-7}\,, & \qquad  
\Mstopsun \lesssim \Mewbh \lesssim  2.9 \times 10^{30}\,~\mbox{GeV}  ~,   
\\
\Mewbh/(2.1 \times 10^{37}~\GeV) \,, &\qquad 2.9 \times 10^{30}~\GeV \lesssim \Mewbh \lesssim 2.8 \times 10^{35}~\GeV ~,
\end{array}
\quad \mbox{(IceCube)}
\right.
\eeqa
where we have chosen $\eta_\nu = \max(1,T_\text{BH}/100~\GeV)$.  The lower limit on the mass in the top line comes from the minimum mass that can be stopped in the Sun. The lower line comes from (\ref{eq:upper-mass-time}), with the upper reach in mass set by (\ref{eq:upper-mass-energy}). If the experiment ran for long enough that it was not limited by (\ref{eq:upper-mass-time}), it would still be limited by the neutrino energy needing to exceed 10 GeV.  In that case, the second line would read $f_{\sbullet} \lesssim 1.8 \times 10^{-6}\,M_{35}^2$ for $2.8\times 10^{34}~\mbox{GeV} \lesssim \Mewbh  \lesssim 2.8\times 10^{35}~\mbox{GeV}$. The above limit can be potentially improved if one takes the temporal information into account. For a heavy PMBH, a transient signal is anticipated and one could search for high-energy solar neutrino flares to search for PMBHs. 

For the Super-Kamiokande searches~\cite{SuperK-2012}, neutrino energy cuts 
$20\,\mbox{MeV} < E_\nu < 55\,\mbox{MeV}$ have been imposed. For the Hawking 
temperature to be in this energy window, $\Mewbh \in (5.1\times 10^{37}, 
1.4\times 10^{38})$~GeV. The observing time is $t_{\rm exp} =2853$~day, so the 
observation-time-related upper limit from \eqref{eq:upper-mass-time} is  
$1.1\times 10^{38}$~GeV, slightly smaller than the energy-related upper mass 
limit. For $\Mewbh < 5.1\times 10^{37}$~GeV, we take the neutrino multiplicity 
factor as $\eta_\nu \sim (T_{\rm BH}/55\,\mbox{MeV})$; otherwise $\eta_\nu =1$. 
Using the experimental upper limit on the flux of 
$183.4\,\mbox{cm}^{-2}\mbox{s}^{-1}$~\cite{SuperK-2012}, we derive the 
following constraints on the PMBH fraction:
\beqa
\label{eq:bound-sun-SK}
\renewcommand{\arraystretch}{1.5}
f_{\sbullet} \lesssim \Bigg \{
\begin{array}{ll}
0.07\,, & \qquad 
\Mstopsun \lesssim \Mewbh \lesssim  5.1\times 10^{37}\,~\mbox{GeV}  ~,   \\
0.03\,M_{38}^{-1} \,, &\qquad 5.1\times 10^{37}\,~\mbox{GeV} \lesssim \Mewbh  \lesssim 1.1\times 10^{38} \,\mbox{GeV} ~. 
\end{array}
\quad \mbox{(Super-K)}
\eeqa
The second line comes from $M_{{\rm max}, t}$ in \eqref{eq:upper-mass-time}. This limit would apply if the drift time were instantaneous, which is a good approximation for $\Mewbh  \lesssim 2\times 10^{38} \,\mbox{GeV}$ from (\ref{eq:tdrift}). Thus, solar neutrino searches cannot probe merging PMBHs to higher masses than this, regardless of experiment time or energy threshold.

\subsection{Baryon number violation process}

Captured PMBHs absorb baryons in the Sun, increasing their mass above 
extremality and leading to $2d$ or $4d$ Hawking radiation depending  on the 
temperature increase by the incoming flux of baryons. For a PMBH captured 
within the Sun, we can approximate the rate of change of the mass, $\Delta M \equiv 
\Mbh - \Mewbh$, as
\beqa
\label{eq:massdecayrate}
\frac{d\Delta M}{dt} = \rho_p\,\pi\,R_\text{EW}^{2}\,v^p_\text{th} - P_{2}\,\Theta(T_\text{BH}-m_{e}) - P_{4}\,\Theta(m_{e}-T_\text{BH}) ~,
\eeqa
where $\Theta(x)$ is the heavyside function and $v^p_\text{th}$ is the proton thermal velocity.
Here, the first term on the right side is energy deposited by the proton flux, 
while the second and third terms are power radiated in $2d$ or $4d$ modes, 
respectively, depending on the temperature. 
As the PMBH absorbs protons, the temperature increases and reaches equilibrium between the first and third terms on the RHS of (\ref{eq:massdecayrate}), giving
\beqa
\label{eq:temp-baryon-equil}
T_{\text{BH}}^\text{eq} = \Big{(}\frac{30\,\rho_{p}\,v^p_\text{th}}{\pi^{2}\, g_{*}} \Big{)}^{1/4}  \, \simeq \, (20\,\mbox{keV}) \times \left( \frac{\rho_p}{50\,\mbox{g}/\mbox{cm}^{-3}} \right)^{1/4} \,\left( \frac{v^p_{\rm th}}{1.7\times 10^{-3}}\right)^{1/4}\,\left( \frac{7.25}{g_*}\right)^{1/4}
\, .
\eeqa
For the Sun, $T^{\rm eq}_{\rm BH} \approx 2 \times 10^{-5}~\text{GeV} \approx 2 \times 10^8~\text{K}$, which is below $m_e$, so the $2d$ radiation is suppressed. 
This gives the total power radiated via \BNV~by surviving PMBHs as
\beqa
L_{\rm \BNV} = N_{\sbullet}^{\text{crit}}P_{4}(T=2\times 10^{-5}~\text{GeV}) = (5.2 \times 10^{15}\,\text{erg} \,\text{s}^{-1} )\,B_{100}^3  ~.
\eeqa
The above power is dramatically smaller than the observed solar luminosity  
$L_\odot=3.8 \times 10^{33}~\text{erg} \,\text{s}^{-1}$. The corresponding neutrino 
flux is also much below the solar neutrino flux~\cite{Bahcall:2004mz}. 
Therefore, proton decay does not constrain the PMBH abundance.

\section{Earth heat and neutrinos}\label{sec:earth} 

As discussed in \cite{Mack:2007xj,Bramante:2019fhi}, if a significant fraction 
of DM is captured as it passes through the Earth and then annihilates 
efficiently to SM particles other than neutrinos inside the Earth, the heat so 
generated would surpass measurements of the internal heat of the Earth.  This 
can be used to set a stringent constraint on DM candidates with scattering 
cross sections that are large enough to be captured by the Earth. 

As discussed in Appendix~\ref{sec:stop-other}, the Earth can efficiently capture PMBHs for $Q \gtrsim \Qstopearth$ or $\Mewbh \gtrsim \Mstopearth$. Similar to \eqref{eq:capture-rate}, the capture rate is estimated to be 
\beqa
\label{eq:capture-rate-earth}
C_{\rm cap} \approx  \epsilon\,\pi \, R_\oplus^2 \, 4\,\pi F_{\sbullet} \approx \left(0.15\,\mbox{s}^{-1} \right) \, \epsilon\,f_{\sbullet}\, M_{26}^{-1}\, ,
\eeqa
where $R_\oplus = 6.4\times 10^8\,\mbox{cm}$. The magnetic field in the Earth core is around 25 gauss~\cite{Earth-core-B}, while the density is around $10\,\mbox{g}/\mbox{cm}^3$~\cite{earth-core-density}. Using \eqref{eq:N-critical}, the critical number to overcome the separation of PMBHs due to the Earth's magnetic field is 
\beqa
N^{\rm crit}_{\sbullet} \simeq (1.5\times 10^{10}) \, M_{26}^{-1} ~. 
\eeqa
The corresponding time to reach this critical number is 
\beqa
\label{eq:tau-crit-earth}
\tau^{\rm crit} = \frac{N^{\rm crit}_{\sbullet}}{C_{\rm cap}} \approx (9.8\times 10^{10}\,\mbox{s})\,f_{\sbullet}^{-1} ~,
\eeqa
which is shorter than the age of the Earth $t_\oplus \approx 1.4\times 
10^{17}$\,s. As in the case of the Sun, if $t_\oplus < \tau^\text{crit}$ then 
PMBHs are separated by Earth's magnetic field, while for $t_\oplus > 
\tau^\text{crit}$ their attractive forces allow for an equilibrium to be 
reached between capture and annihilation. 
The drift time for the Earth is very similar to that of the Sun in (\ref{eq:tdrift}) because the average densities and $v_\text{th}$ (for Sun)/$v_{\rm F}$ (for Earth) are similar to within $\cO(1)$.  Specifically, using $n_e = 1.7 \times 10^{24}~\cm^{-3}$ and Fermi energy $E_{\rm F}=m_e v_{\rm F}^2/2=1~\text{eV}$ (see Appendix \ref{sec:stop-other}), $t_\text{drift} \sim (1 \times 10^6~\text{s}) M_{26}$. We will conservatively require $t_\text{drift} < t_\oplus$ for annihilations to become important, setting an upper bound on the mass that can be probed.  Thus, PMBHs with $\Mewbh \lesssim 1 \times 10^{37}~\GeV$ will reach equilibrium in the Earth's center.

If equilibrium is reached, the time-averaged power generated by PMBH annihilations is independent of $\Mewbh$ and is estimated to be
\beqa
P_A \simeq (2.4\times 10^{15}\,\mbox{W})\,f_{\sbullet} ~. 
\eeqa
Compared to the internal heat of the Earth $P_\oplus \approx 4.7\times 10^{13}$~W~\cite{earth-heat}, this sets a bound
\beqa
\label{eq:bound-earth-heat}
&&f_{\sbullet} \lesssim 0.02 \qquad \qquad \qquad  (\mbox{Earth heat}) ~,
\eeqa
again for $\Mstopearth\,\lesssim \,\Mewbh \lesssim 1 \times 10^{37}~\GeV$. 
However, unlike for particle DM, the rate of mergers must be considered for 
heavy PMBHs.  As a rough estimate, to set a bound we require the annihilation 
rate (which equals the capture rate in equilibrium) to be faster than the 
diffusion rate of heat from the core to the surface. The thermal diffusion 
timescale of the inner core alone is estimated as $\tau_\kappa =(1.4 \pm 0.7) 
\times 10^9~\text{yr} = (4.4 \pm 2.2) \times 10^{16}~\text{s}$ 
\cite{10.1111/j.1365-246X.2011.05222.x}, which is comparable to the age of the 
Earth.  Demanding $C_\text{cap} \gtrsim \tau_\kappa^{-1}$ requires $\Mewbh 
\lesssim 7 \times 10^{41} \, f_{\sbullet}$, comparable to the bound from 
$t_\text{drift} < t_\oplus$ when $f_{\sbullet}=1$. Thus, we do not expect a 
dramatic reduction in sensitivity even when the PMBH merger/annihilation events 
are very rare.  In fact, this long diffusion timescale enables us to set bounds 
for much larger masses, rather than being limited by how long we have been 
taking detailed measurements of Earth's heat flux.

In addition to Earth heating, the neutrino flux from PMBH mergers can be used to set a limit.  We recast the results in the IceCube search for particulate DM $\chi$ \cite{Aartsen:2016fep}.  Specifically, they set a mass-dependent bound on the rate of $\chi$ particle annihilations $\Gamma_{A,\chi}$ going to both hard and soft channels.  We expect post-merger PMBHs with a temperature $T_\text{BH}$ given by (\ref{eq:TBH-merger}) will emit neutrinos with a similar energy spectrum as particulate DM whose mass $m_\chi \sim T_\text{BH}$ in both the hard ($\chi \chi \to W^{+} W^{-}$) and soft ($\chi \chi \to b \bar{b}$) channels.  The only difference is that the neutrino multiplicity is enhanced by a factor $\sim \eta_\nu \Mewbh/T_\text{BH}$.  As before, $\eta_\nu$ is a neutrino multiplicity parameter, with $\eta_\nu \sim 1$ for the hard component and $\eta_\nu \gg 1$ for the soft component.  Thus, their bound on $\Gamma_{A,\chi}$ for a given $m_\chi$ becomes a bound on $\Gamma_A\,\eta_\nu \Mewbh/T_\text{BH} \lesssim \Gamma_{A,\chi}$ for a given $\Mewbh$ satisfying $T_\text{BH} \geq m_\chi$ in (\ref{eq:TBH-merger}).  When captures and annihilations are in equilibrium, $\Gamma_A=C_\text{cap}/2$, with $C_\text{cap}$ given in (\ref{eq:capture-rate-earth}).

IceCube sets a bound on DM particles $\chi$ annihilating to neutrinos in the Earth with masses $10~\GeV<m_\chi<10^4~\text{GeV}$ \cite{Aartsen:2016fep}.  Post-merger PMBHs are expected to emit a hard component of neutrinos with energy comparable to their temperature.  Using (\ref{eq:TBH-merger}), $T_\text{BH}$ is in the range of $m_\chi$ probed by IceCube when $3 \times 10^{32}~\GeV \lesssim \Mewbh \lesssim 3 \times 10^{36}~\GeV$.  However, the IceCube search runs only 327 days.  To have at least one annihilation during this time, assuming both equilibrium and $\epsilon=1$, $\Mewbh \lesssim (2 \times 10^{32}~\GeV) f_{\sbullet} (t_\text{exp}/327~\text{day})$ is required [analogous to \eqref{eq:upper-mass-time}].

At smaller masses, the post-merger temperature is higher.  Thus, one could use IceCube to search for higher-energy neutrinos above the maximum mass of 10 TeV used in the DM search. However, the Earth is opaque to neutrinos with energies higher than 40 TeV \cite{Aartsen:2017kpd}.  Thus, such a search would be ineffective.

Alternatively, one could try to estimate the soft component of neutrinos coming from these lower-mass, higher-temperature PMBH mergers.  Using a similar approximation as the solar case, we may expect a neutrino multiplicity enhancement of $\eta_\nu \sim T_\text{BH}/ E_\nu$, with the neutrino energy $E_\nu \lesssim 40~\TeV \lesssim T_\text{BH}$.  Then, the average rate of neutrino production at energy $E_\nu$ is
\begin{equation}
\Gamma_A \, \eta_\nu \Mewbh/T_\text{BH} \sim C_\text{cap} \Mewbh/E_\nu \approx (1.5 \times 10^{21}~\text{s}^{-1}) \epsilon f_{\sbullet} (E_\nu/10~\TeV)^{-1} ~,
\end{equation}
independent of mass in these approximations.  This quantity should be compared against the IceCube bound $\Gamma_{A,\chi} < 1.47 \times 10^{10}~\text{s}^{-1}$ for $m_\chi=10~\TeV$ in the soft channel \cite{Aartsen:2016fep}. 
Thus, there is an approximate bound $f_{\sbullet} \lesssim 10^{-11}$ for 
$\Mewbh \lesssim (2 \times 10^{32}~\GeV) f_{\sbullet}$ [from 
\eqref{eq:upper-mass-time} and using the Earth capture rate in 
\eqref{eq:capture-rate-earth}].  The exact bound on $f_{\sbullet}$ depends 
sensitively on the details of the soft neutrino multiplicity, which requires 
detailed numerical study of the Hawking radiation and its interactions and 
showering within the Earth, beyond the scope of this work. However, this 
assumed an instantaneous $\tau^\text{crit}$, neglecting Earth's magnetic 
field.  Taking the value in (\ref{eq:tau-crit-earth}) into account, the Earth 
neutrino bound is reduced to
\beqa
\label{eq:bound-earth-IC}
\renewcommand{\arraystretch}{1.5}
f_{\sbullet} \lesssim \Bigg \{
\begin{array}{ll}
7 \times 10^{-7}\,, & \quad 
\Mstopearth \lesssim \Mewbh \lesssim  1.4\times 10^{26}\,~\mbox{GeV}  ~,   \\
\Mewbh/(2 \times 10^{32}~\GeV) \,, &\quad 1.4 \times 10^{26}~\GeV \lesssim \Mewbh \lesssim 3 \times 10^{36}~\GeV ~. 
\end{array}  
~ \mbox{(IceCube)}
\eeqa
The low-mass cutoff is from the minimum stopping mass in the Earth, and the high-mass behavior is from requiring at least one annihilation during the observation period of the search.  Because this is much weaker than the maximum bound with instantaneous $\tau^\text{crit}$, we expect it to be fairly robust against the neutrino multiplicity approximations we employed.

Earth heating and neutrino production can also occur from the \BNV~process.   
The temperature of PMBHs due to proton absorption can be calculated using (\ref{eq:temp-baryon-equil}).
With $v^p_\text{th}=(3\,T_\text{core}/m_{p})^{1/2}$, $\rho_p = 12.2\,\mbox{g}/\mbox{cm}^{3}$ and $T_\text{core}\approx 5700$~K, $T_{\text{BH}}^\text{eq}=5.4~\text{keV}=6.3 \times 10^{7}~\text{K}$. We can evaluate the power radiated by the proton decay as 
\begin{equation}
L_{\rm \BNV} = N_{\sbullet}^{\text{crit}}P_{4}(T=5.4~\text{keV}) = 1.2 \times 10^{6} \, \text{W} ~,
\end{equation}
which is less than the internal heat of Earth and thus does not give any constraints.
There are no direct experimental measurements for geoneutrinos at this low energy.

\section{Neutron stars and white dwarfs}\label{sec:neutron-star} 

\subsection{PMBH capture}

PMBHs can get captured by other large bodies outside the 
solar system, such as 
neutron stars and white dwarfs. The calculation proceeds 
similarly to solar 
capture, though in the case of neutron stars, we include a 
relativistic 
correction factor,
\begin{equation}
\label{eq:Ccap-NS}
C_{\rm cap}\, \approx\, \epsilon \, \pi \, R^2 \, 
\left[\frac{1+ (v_{\rm 
		esc}/v)^2}{1 - v_{\rm esc}^2}\right] \, 4 \, \pi 
F_{\sbullet} \approx (0.11\,\mbox{s}^{-1})\,f_{\sbullet} \,R_{10}^2\,M_{26}^{-1}~,
\end{equation}
where $R=R_{10}\times 10$~km is the radius of the neutron star, $v_{\rm esc}\approx 0.47$ for a neutron star and $v=10^{-3}$. In Appendix \ref{sec:stopping}, we show that a MeBH will be stopped if 
it enters a neutron star or white dwarf, so $\epsilon=1$.
The number of captured PMBHs per neutron star is then 
\begin{equation}\label{eq:nucleon-capt}
N_{\sbullet}^{\rm NS} = C_{\rm cap}\, \tau_\text{NS} \sim \left(3.3 \times 10^{16}\right)~ f_{\sbullet} 
\,R_{10}^2\, M_{26}^{-1}  \, 
\tau_{10} ~,
\end{equation}
where $\tau_\text{NS} = \tau_{10} \times 10^{10}~{\rm yr}$ is the age of the neutron star.  The 
number of captured PMBHs per white dwarf (with $R \approx 7000~\text{km}$ and $v_\text{esc} \approx 0.02$) is 
\begin{equation}
N_{\sbullet}^{\rm WD} \sim \left(2.3 \times 10^{19}\right)~f_{\sbullet} \, 
M_{26}^{-1} \, 
\tau_{10} ~.
\end{equation}

Note that in the above analysis, we neglected the strong magnetic field of the 
neutron star. The $B$ field of old neutron star is less than $10^8~{\rm 
	gauss}$, which induces a force of order $F_{\rm EM} \simeq Q\,h \,B$. The gravitational force close to the surface of the 
neutron star is of order $F_{\rm grav}\simeq G \, M_\text{NS} \, \Mewbh / R^2$.  Neglecting relativistic corrections, which are $\mathcal{O}(1)$, for simplicity, the ratio of the forces is
\beqa
\frac{F_{\rm EM}}{F_{\rm grav}} \, \sim \, \frac{2\sqrt{\pi} \, B \, M_{\rm pl} \, 
	R^2}{c_W\, M_\text{NS}} \sim (7.8 \times 10^{-10}) \,\times\, \frac{M_\odot}{M_\text{NS}} \, 
	R_{10}^{2} \,  
B_8 ~,
\eeqa
for a surface magnetic field $B = B_8 \times 10^{8}~{\rm gauss}$, the approximate value for an old neutron star.
Old neutron stars are more prevalent and will have captured many more PMBHs than young neutron stars, so they are the most relevant for setting bounds. The magnetic force is much smaller than the gravitational force and can therefore be neglected to first approximation for the calculation of the encounter rate.

\subsection{PMBH distribution}

Like the Sun, magnetic fields inside neutron stars can separate PMBHs and prevent them from merging.  The analysis of this possibility is complicated in this scenario by the unknown exotic phase of matter in the core of the neutron star.  We begin by analyzing that case.

At a depth of order few km below the crust, a neutron star is expected to be a 
proton superconductor~\cite{1959NucPh..13..655M,1969Natur.224..872B} 
(see Ref.~\cite{Gezerlis:2014efa} for a recent review), in which protons form Cooper 
pairs. This region forms the outer core.  Magnetic fields in this region are 
confined to flux tubes at early times, which diffuse outward until the core is 
free of magnetic field. The diffusion time is somewhat uncertain, but can be 
less than the typical age of an old neutron star~\cite{Harvey:1985nv}.

The nature of the inner core is not definitively known. For example, the 
environment could be a pion superconductor in which the pions acquire 
an isospin-breaking expectation value~\cite{PhysRevD.12.979}. It could also be 
a color superconductor, which would break $SU(3)_c \times U(1)_{\rm EM} \to 
SU(2) \times U(1)$, with the unbroken $U(1)$~\cite{Alford:2001dt}.

We are 
most concerned, as we will see, with the phase at a distance of around a kilometer from the center, 
very deep in the neutron star.  There are essentially two qualitatively different scenarios.  If the neutron star is a proton or pion superconductor at a distance of around a kilometer from the center, then the magnetic field lines of the PMBHs are confined to quantized flux tubes which create a large outward force that 
can balance the gravitational pull toward the core of the neutron star. 
Otherwise, if the magnetic field lines are not confined to flux tubes, there is a weak outward magnetic force (similar to the case of the Sun in Fig.~\ref{fig:sun-separation}), and the PMBH and anti-PMBHs collapse 
unimpeded toward the core.

As a PMBH enters a superconductor, its magnetic field is confined 
to flux tubes. In the case of a proton superconductor, the flux of each tube is
\begin{equation}
\Phi = \frac{\pi}{e} ~,
\end{equation}
which is half of the fundamental Dirac charge. The factor of half is due to the fact 
that it is charge-two Cooper pairs of protons that break the electromagnetic gauge group. 
For a pion superconductor, where EM charge is broken only by one unit, the flux 
per tube is twice that. There must then be $2 \, Q$ ($Q$) flux tubes per PMBH for a proton (pion) 
superconductor. For simplicity, we focus on the proton superconductor case, 
though the pion superconductor case is not qualitatively different.
The typical size of 
these flux tubes is given by the London penetration depth
\begin{equation}
\lambda = \left(\frac{m_p}{e^2 \, n_p}\right)^{1/2} \sim 10^{-12}~{\rm cm} ~,
\end{equation}
for $n_p m_p \approx 4\times 10^{14}\,\mbox{g}/\mbox{cm}^3$. The typical magnetic field inside the flux tube is then given by
\begin{equation}
B_\Phi \sim \frac{\Phi}{\pi \, \lambda^2} \sim 10^{16}~{\rm gauss} ~,
\end{equation}
much larger than the surface magnetic field.

\begin{figure}[t]
	\includegraphics{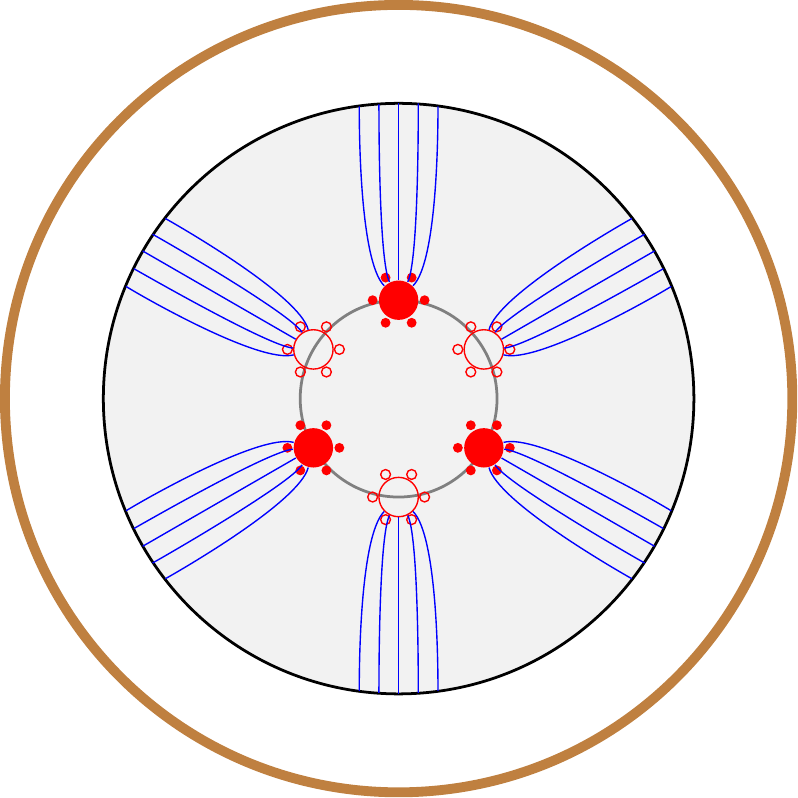}
	\caption{An ensemble of PMBHs (filled red) and anti-PMBHs (hollow red) 
	inside the superconducting core of a neutron star. The gray circle is the stable  
	position where the magnetic and gravitational forces are equal.  The 
	black circle is the edge of the superconducting region. The thick brown 
	circle indicates the edge of the neutron star. The blue lines 
	represent magnetic flux tubes. Magnetic field lines are not depicted outside the superconducting core.  For color superconductors or other phases where magnetic fields do not confine in flux tubes, the dynamics are more similar to Fig.~\ref{fig:sun-separation}.
	}
		\label{fig:fluxtubes}
\end{figure}
A flux tube has an enormous tension force that would like to minimize the 
length of the flux tube. This force is given by~\cite{Harvey:1985nv}
\beqa
F_{\rm T} \sim B_\Phi^2\,\pi\,\lambda^2\, \ln{\left( \lambda/\xi\right)}\sim 
10^4~{\rm N} ~,
\eeqa
where $\xi \sim {\rm few} \times 10^{-13}~{\rm cm}$ is the Bardeen-Cooper-Schrieffer (BCS) correlation 
length for each flux tube. For GUT monopoles, the tension force is sufficient to eject 
monopoles inside the superconducting core when it forms or to prevent monopoles 
from entering except along flux tubes~\cite{Harvey:1985nv}. For PMBHs, however, the charge-to-mass 
ratio is lower and the gravitational force allows the PMBHs to penetrate into 
the superconducting region. The flux tubes are approximately radially outward 
going from the entry point of the PMBH. This configuration is qualitatively 
illustrated in Fig.~\ref{fig:fluxtubes}. For a sufficient tension force, this 
allows for a stable shell of hanging PMBHs, also illustrated in 
Fig.~\ref{fig:fluxtubes}. The radius of the shell is given by balancing the 
tension force with the neutron star gravitational force, neglecting for the moment 
self-gravitation contributions of the PMBH population,
\begin{equation}
R_\text{balance} \approx \frac{6 \, Q\, F_{\rm T}}{4 \, \pi \, G \, \rho_c \, \Mewbh} \sim 1600~{\rm m}  ~,
\end{equation}
with $\rho_c \approx m_p\, n_p\approx 4\times 10^{14}\,\mbox{g}/\mbox{cm}^3$. 

The energy of the flux tubes can be lowered if outward-going flux tubes from 
PMBHs merge with inward going flux tubes from anti-PMBHs, leading to a tension 
force that pulls the PMBH and anti-PMBH together to annihilate. The dominant 
effect that initially prevents this from happening is that the drift of the 
flux tubes is very slow.~\footnote{There is also a potential barrier given by 
the added tension of a configuration where the two flux tubes are merged rather 
than radially outward, but this potential will be negligible in the regime of 
small separation we consider.} The drift velocity can be obtained by balancing 
the flux tube tension against the force due to impinging electrons and other 
elements of the degenerate fluid in the neutron star (see Refs.~\cite{Harvey:1985nv,Jones:2006} for estimation of the drift velocity).

The flux tubes from a given PMBH are grouped into ``bundles'' stretching roughly vertically from the PMBH (see Fig.~\ref{fig:fluxtubes} for an illustration). The spread of the flux tubes within a bundle is set by balancing the tension that restores the flux tubes toward the vertical with the pressure due to collisions of electrons in the plasma with the flux tube. The angular scale for this spread can be obtained by comparing the energy required to displace the flux tube from vertical by an angle $\theta$ with the temperature. 
In the limit of $\theta \ll 1$, the ratio is $F_{\rm T} \, R_c \,\,\theta^2/(2 \, T) \sim 1$
for $T \simeq 10^6\,\mbox{K}$ and $R_{c10}=R_c/(10~\text{km})$ the superconducting core radius in units of 10 km.  Additionally, the flux tubes will not pack more tightly than $\sim \lambda$~\cite{Zhao-2017}, so $\theta_{\rm T} \gtrsim \sqrt{Q} \lambda/R_c$.  Putting these together, 
\begin{equation}
\theta_{\rm T} \simeq \max \left[(5.3\times 10^{-13})\,R_{c10}^{-1/2} \, , ~ (1.3 \times 10^{-15}) M_{26}^{1/2} R_{c10}^{-1} \right] ~.
\end{equation}
When the separation angular distance from one bundle to another bundle is larger than $\theta_{\rm T}$, the tension force is more important and one can treat the whole bundle as one composite object; otherwise, the thermal pressure is more important. This condition can be translated into
\beqa
\label{eq:bundle-condition}
N_{\sbullet} \lesssim  4\pi\,\theta_{\rm T}^{-2} \approx \mbox{min}[  (4.5 \times 10^{25})\,R_{c10},  (7.4 \times 10^{30}) \,R_{c10}^2\,M_{26}^{-1} ] ~. 
\eeqa

We can estimate the total encounter rate of two bundles using $\Gamma_{\rm enc} = N_{\sbullet}\, n_2\, \sigma \, v_{\sbullet}$, with the $2d$ number density as $n_2 \simeq N_{\sbullet}/(4\pi R_c^2)$, $\sigma \simeq  R_c\,\theta_{\rm T}$, and $v_{\sbullet} \simeq \sqrt{T/\Mewbh}$. For two bundles that cross, we can estimate the merger or annihilation probability by requiring $\mathcal{O}(Q)$ pairs of flux tubes merge. This probability is estimated to be $\mathcal{P} \simeq \mbox{min}[(Q\,\lambda\,v_{\rm ft})/(R_c\,\theta_{\rm T}\,v_{\sbullet}), ~ 1]$ with $v_{\rm ft} \simeq \sqrt{T/(F_{\rm T} R_c)}$ as the flux tube thermal velocity. Numerically, one has $\mathcal{P} \simeq \mbox{min}[(4.7\times 10^4) \,M_{26}^{3/2}R_{c10}^{-1}, ~(1.5 \times 10^7) \,M_{26}R_{c10}^{-1/2}, ~ 1]$, which is 100\% for $\Mewbh \gtrsim (7.6\times 10^{22} \,\mbox{GeV})R_{c10}^{2/3}$ (the second term from when the tubes are tightly packed is never relevant for $\mathcal{P}$). The merger or annihilation rate is 
\begin{equation}
\Gamma_A = \Gamma_{\rm enc}\, \mathcal{P}\, = 
\mbox{max} \left[ (3.7 \times 10^{-26}\,\mbox{s}^{-1}) \,M_{26}^{-1/2}\,R_{c10}^{-3/2} \, , ~ (9.1 \times 10^{-29}\,\mbox{s}^{-1})\,R_{c10}^{-2} \right] \,\mathcal{P}\,N_{\sbullet}^2 ~.
\end{equation}
The equilibrium time for the annihilation and capture rates is
\begin{equation}
\tau_{\rm eq} \approx \mbox{min} \left[ (3.6\,\times 10^5\,\mbox{yr}) \,M_{26}^{3/4}\, R_{c10}^{3/4} \, , ~ (7.2 \,\times 10^6\,\mbox{yr}) \,M_{26}^{1/2}\, R_{c10} \right] \, f_{\sbullet}^{-1/2} \, R_{10}^{-1}\,\mathcal{P}^{-1/2} ~,
\end{equation}
which is shorter than the age of neutron stars for $\Mewbh \lesssim (8.5 \times 10^{31}\,\mbox{GeV})\,f_{\sbullet}^{2/3}\, R_{c10}^{-1} \, R_{10}^{4/3} \tau_{10}^{4/3}$.  The second term inside the minimum is never important when comparing $\tau_\text{eq}$ to the age of neutron stars $\tau_{\rm NS} \sim 10^{10}~\text{yr}$, so we can henceforth neglect it. The number of PMBHs in a neutron star is
\beqa
\label{eq:number-ns-eq}
N_{\sbullet}^{\rm cap} &=& \mbox{min}[\tau_{\rm eq}, \tau_{\rm NS}] \,C_{\rm cap} \nonumber \\ 
& \approx& \mbox{min}[ (1.2 \times 10^{12}) \,  f_{\sbullet}^{1/2}\,M_{26}^{-1/4}\, R_{c10}^{3/4} \, R_{10} \,\mathcal{P}^{-1/2} , ~ (3.3 \times 10^{16})~ f_{\sbullet} 
\,R_{10}^2\,M_{26}^{-1} \tau_{10}  ]  ~. 
\eeqa
Note that this number of $N_{\sbullet}$ satisfies the condition in \eqref{eq:bundle-condition}.

One might also worry that the increased number of PMBHs will cause the stable 
shell of ``hanging'' PMBHs to shrink and destabilize due to the 
self-gravitation of the PMBHs. This happens when the flux tube tension is not 
sufficient to balance the combined radial gravitational force of the PMBHs and 
the neutron star. This occurs for a critical number $(9\, e^3 \, F_{\rm T}^3  \, M_{\rm pl}^3)/(2\,c_W^3 \, \pi^{7/2} \Mewbh \, \rho_c^2) \sim 10^{29}\, M_{26}^{-1}$, which is larger than the maximal captured number for the whole neutron star age. Therefore, this effect is also negligible.

If the inner core has unconfined magnetic fields, as is the case for a color 
superconducting phase, then the magnetic fields are drastically weaker, of 
order the surface magnetic field.  This is analogous to monopole separation in the uniform field of the Sun, see Fig.~\ref{fig:sun-separation}. The magnetic fields are then insufficient to 
sustain a stable separation of PMBHs and anti-PMBHs, and annihilation proceeds 
uninhibited until it equilibrates with the capture rate once a critical number of captured 
PMBHs is reached. The critical number is given by \eqref{eq:N-critical},
\begin{equation}
\label{eq:Ncrit-NS-unconfined}
N^{\rm crit}_{\sbullet}  = (6 \times 10^2) \, M_{26}^{-1} \, B_8^3 ~.
\end{equation}
The annihilation rate is given 
by \eqref{eq:annihilation-rate} and \eqref{eq:Ccap-NS}, 
\begin{equation}
\Gamma_A = (0.05~{\rm s}^{-1})\times f_{\sbullet} \, M_{26}^{-1} ~.
\end{equation}

For a white dwarf, the situation is simpler as they are not expected to be 
superconducting. 
Similar to the case for the Sun or a color superconducting neutron star core, annihilation quickly 
equilibrates with capture at $N_{\sbullet}^{\rm crit}$, with 
\begin{equation}
N^{\rm crit}_{\sbullet}  = 10^{14} \, M_{26}^{-1} \, B_6^3 ~,
\end{equation}
taking the white dwarf magnetic field to be $B_6 \times 10^6~{\rm gauss}$ and the density to 
be $\rho_c\sim 10^6~{\rm g}/{\rm cm}^3$. The annihilation rate is then
\begin{equation}
\Gamma_A = (37~{\rm s}^{-1})\times f_{\sbullet} \, M_{26}^{-1} ~.
\end{equation}

\subsection{Constraints}

In the scenarios that we consider, there are two potential sources of heat generation in 
the star: annihilation and \BNV.  The luminosity generated by 
annihilation which has 
equilibrated  with capture is given simply by
\begin{equation}
L_A = \Mewbh \, C_{\rm cap} ~.
\end{equation}
\BNV~has a cross section that is at least 
given by the EW 
radius $R_{\rm EW}$ as argued above. Thus, the luminosity 
generated by nucleon decay 
is given by
\begin{equation}
L_{\rm \BNV} \approx N_{\sbullet}^{\rm cap} \, \rho_c \, \pi \, R_{\rm 
	EW}^2 \, \langle v 
\rangle ~,
\end{equation}
where $\langle v \rangle$ is the mean speed of nucleons in 
the core of the 
star.  For neutron stars, this is given by $\langle v 
\rangle \approx 3 \, v_{\rm F} 
/ 4$, where $v_{\rm F} \sim 0.2$ is the Fermi velocity. For white 
dwarfs, the protons 
at the core are not expected to be degenerate, so the 
average velocity is set 
by the temperature, $\langle v \rangle = 2 \sqrt{2 \, T/(\pi 
	\, m_p)} \approx 5\times 10^{-4}$, with $T 
/ m_p \sim  10^{-7}$.  

We now consider three scenarios: a neutron star with magnetic fields confined to flux tubes in the core, a neutron star without flux tubes, and a white dwarf. In the first 
case, the dominant luminosity comes from \BNV. Using the abundance of captured PMBHs in \eqref{eq:number-ns-eq}, the luminosity due to \BNV~is 
\beqa
L^{\rm NS}_{\rm \BNV} = \mbox{min}[(1.2 \times 10^{32}~{\rm erg} \, \mbox{s}^{-1})\,\mathcal{P}^{-1/2} f_{\sbullet}^{1/2}\,M_{26}^{3/4}\, R_{c10}^{3/4} \, R_{10} , ~ (3.3 \times 10^{36}~{\rm erg} \, \mbox{s}^{-1})\, f_{\sbullet}\,R_{10}^{2} \tau_{10}  ] ~.
\eeqa
The luminosity can be compared either to that of individual stars, the total from all stars, or to diffuse emissions. There are advantages and disadvantages to each approach as outlined in Ref.~\cite{Kolb:1984yw}. All of these bounds are on the order of $L_\gamma < 10^{32}~{\rm erg} \,{\rm s}^{-1}$ per neutron star~\cite{Kolb:1982si,Freese:1983hz,Kolb:1984yw}.  
Assuming a fraction $r = L_{\rm tot} / L_\gamma$ of the total luminosity to photon luminosity, a bound is obtained after expanding $\mathcal{P}$: 
\beqa
\label{eq:bound-ns-ind}
\renewcommand{\arraystretch}{1.5}
f_{\sbullet} \lesssim \left\{
\begin{array}{ll}
3.2 \times 10^4 \, R_{c10}^{-5/2} \, R_{10}^2 \, r^2 \, , & 
 \frac{\Mewbh}{\GeV}  \lesssim (7.6 \times 10^{22}) R_{c10}^{2/3} \,,
\\
0.69\,M_{26}^{-3/2}\, R_{c10}^{-3/2} \, R_{10}^{-2} \, r^2 \,, &
(7.6 \times 10^{22}) R_{c10}^{2/3} \lesssim \frac{\Mewbh}{\GeV}  \lesssim (8 \times 10^{28}) R_{c10}^{-1} r^{2/3} \tau_{10}^{2/3}  ~,   
\\
(3 \times 10^{-5})\,R_{10}^{-2}\, \tau_{10}^{-1} \, r \,, & 
(8 \times 10^{28}) R_{c10}^{-1} r^{2/3} \tau_{10}^{2/3} \lesssim  \frac{\Mewbh}{\GeV}    \lesssim  (9.9 \times 10^{45}) r~, 
\\
\frac{\Mewbh}{(3.3 \times  10^{50}\,\GeV)}\, R_{10}^{-2}\, \tau_{10}^{-1}  \,, &  (9.9 \times 10^{45}) \lesssim   \frac{\Mewbh}{\GeV}   \,.
\end{array}  
\right.
(\mbox{\BNV})
\eeqa
Here, the limit in the last line comes from requiring at least one captured PMBH for all $\sim 10^8$ neutron stars in our galaxy [see \eqref{eq:nucleon-capt}], which is valid for the diffuse emission limits.
For proton superconductors, $r\sim 
1$~\cite{TSURUTA1979237,1981ApJ...244L..13V,Kolb:1984yw}. If the core is 
instead a pion condensate, $r$ could be around $10^3$ or 
$10^4$~\cite{TSURUTA1979237,1981ApJ...244L..13V,Kolb:1984yw}, leading to a 
reduced bound.

If the core is not superconducing or if it is a color superconductor, then 
annihilation equilibrates with capture at a lower captured PMBH abundance. The luminosity from annihilation is still subdominant compared to the luminosity from \BNV, which is [using (\ref{eq:Ncrit-NS-unconfined})]
\begin{equation}
L^{\rm NS}_{\rm \BNV} =  \left(1 \times 10^{23}~{\rm erg} \,\mbox{s}^{-1}\right)\, f_{\sbullet} \,B_8^3~.
\end{equation}
Although \BNV~dominates annihilation luminosity, it is still unlikely to contribute a relevant bound, 
even if all of the power generated is emitted as photons. 
\footnote{The kinetic energy of captured PMBHs can also heat up neutron stars~\cite{Baryakhtar:2017dbj}. Given the semi-relativistic velocity of the captured PMBHs, the kinetic heating is subdominant to the annihilation-produced luminosity.}

For white dwarfs, the constraint determination proceeds similarly.  The 
luminosity due to 
annihilation is given by $L^{\rm WD}_{\rm ann} = \left(1.6 \times 10^{25}~{\rm erg}\, \mbox{s}^{-1}\right) \, f_{\sbullet}$,
while the luminosity due to \BNV~is given by $L^{\rm WD}_{\rm \BNV} = \left(3.3 \times 10^{23}~{\rm erg}\, \mbox{s}^{-1}\right) \, f_{\sbullet}$. The power from the dimmest nearby white dwarfs is around $2 \times 10^{29}~{\rm erg}\, {\rm s}^{-1}$~\cite{Giammichele_2012}, which is a few orders of magnitude larger than that 
generated by PMBHs such that no relevant constraint is set.

\section{Discussion and conclusions}
\label{sec:conclusion}

\begin{figure}[t!]
	\label{fig:bounds}
	\begin{center}
		\hspace{-0.5cm}\includegraphics[width=0.65\textwidth]{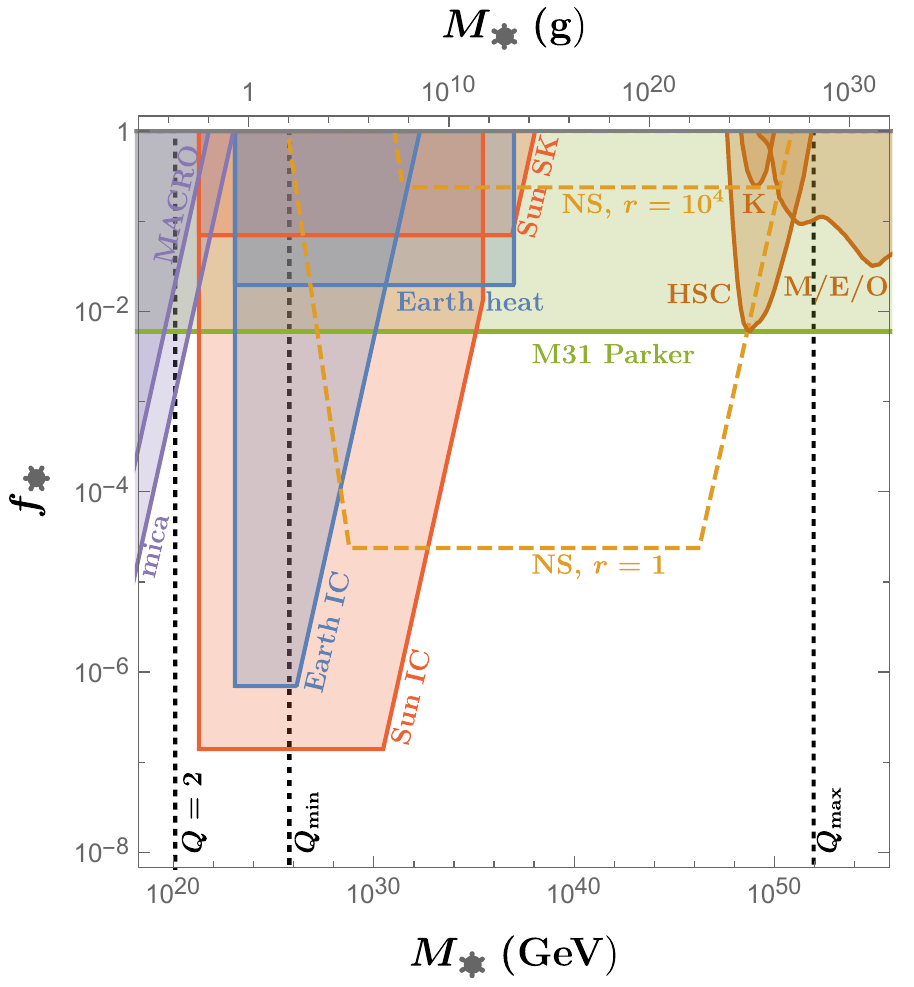}
		\caption{Bounds on PMBH abundance as a fraction of the dark matter abundance.  In green is the Parker bound using M31/Andromeda.  Red and blue show constraints from the Sun and the Earth, respectively, due to neutrino observations at IceCube (IC), Super-Kamiokande (SK), and Earth heating.  Orange dashed lines show constraints from neutron stars (NS) assuming a total baryon number violation energy on emitted photon luminosity of either $r=1$ or $r=10^4$.  See details and caveats in the text. Purple regions are excluded by direct searches from MACRO and ancient mica.  Brown displays constrains from microlensing at Subaru/HSC (HSC), Kepler (K), and MACHO/EROS/OGLE (M/E/O).  The dotted black vertical lines show where $Q=2$, $Q_\text{min}\simeq 10^6$ (assuming the existence of a GUT monopole), and $Q_\text{max}\simeq 1.4\times 10^{32}$ (above which there is no EWS corona).}
	\end{center}
\end{figure}

The constraints on PMBHs are summarized in Fig.~\ref{fig:bounds}.  
The Parker bound due to the coherent magnetic fields in M31/Andromeda comes from Eq.~(\ref{eq:parker-limit-M31}).  The bounds from solar neutrinos come from Eq.~(\ref{eq:bound-sun-IC}) for IceCube (IC) and (\ref{eq:bound-sun-SK}) for Super-K (SK).  The Earth neutrino bound from IceCube is in (\ref{eq:bound-earth-IC}), while the Earth heating bound is (\ref{eq:bound-earth-heat}). The merger bounds are limited on the left by requiring the stopping length be smaller than the size of the bodies, and on the right by requiring at least one merger during the relevant timescale for the bound.
We have approximated the left boundaries as sudden cutoffs at $Q_\text{stop,min}$ in Table \ref{tab:quantities}, but a more thorough investigation could account for the nonzero probability to capture smaller-charge PMBHs.
The bounds from \BNV~processes in neutron stars are shown assuming either $r=1$ (for a proton superconductor) or $r=10^4$ (for a pion superconductor), setting $R_{10}=1$ and $R_{c10}=0.7$ in (\ref{eq:bound-ns-ind}).
The left-most portion of the bounds becomes weaker at smaller masses because PMBHs annihilate.  In the center, where the bound is a horizontal line, PMBHs do not annihilate efficiently.  To the right, as the mass increases some neutron stars do not capture any PMBHs, decreasing the \BNV~luminosity.
If neutron star cores do not confine magnetic fields to flux tubes, such as if they have color superconducting cores, neutron stars provide no bound.  Thus, this bound is quite tentative, depending on the details of neutron star cores.
Mass and charge are related by $\Mewbh/Q = c_W\sqrt{\pi}M_{\rm pl}/e \approx 5.1\,M_{\rm pl}$, with $Q_\text{min}$ from above Eq.~(\ref{eq:mass-range}), which assumes the existence of a GUT monopole with mass $\Mmono=10^{17}~\GeV$, and $Q_\text{max}$ from (\ref{eq:Qmax}), above which there is no EWS corona.

At higher masses, PMBHs can be constrained by gravitational lensing of stars.  Subaru/HSC set limits using stars in Andromeda \cite{Niikura:2017zjd}  for BH masses starting from $\Mewbh \gtrsim 10^{23}~\text{g}$ \cite{Smyth:2019whb}. Other microlensing studies from Kepler \cite{Griest:2013aaa}, MACHO \cite{Allsman:2000kg}, EROS \cite{Tisserand:2006zx}, and OGLE \cite{Wyrzykowski:2011tr} cover the remainder of the possible PMBH masses up to $Q_\text{max}$.  These bounds will not be sensitive to the presence of the EWS corona \cite{Croon:2020wpr,Bai:2020jfm}, especially because the corona is inside the Einstein radius.  Eventually, microlensing of X-ray pulsars \cite{Bai:2018bej} and femtolensing \cite{Katz:2018zrn} or lensing parallax \cite{Jung:2019fcs} of gamma ray bursts  could set bounds down to $\Mewbh \gtrsim 10^{17}~\text{g}$.

The MACRO experiment sets a flux limit $F < 1.6 \times 10^{-16}~\cm^{-2}~\text{s}^{-1}~\text{sr}^{-1}$ on heavy monopoles ($\Mmono \geq 10^{17}~\GeV$) with $v = 10^{-3}$ \cite{Ambrosio:2002qq}.  Comparing to (\ref{eq:local-flux}), we obtain a na\"ive bound of $f_{\sbullet} < 1.7 \times 10^{4}\,M_{26}$. We do not expect the bound to change very much for $Q>1$ or if the PMBH also contains some residual electric charge \cite{Ambrosio:2002qq,Lehmann:2019zgt}. Thus, MACRO sets constraints for $\Mewbh < 6 \times 10^{21}$~GeV. Similarly, searches for tracks in ancient mica samples on monopoles with $v=10^{-3}$ give a constraint between $F < 10^{-17}\,\cm^{-2}\,\text{s}^{-1}\,\text{sr}^{-1}$ to $3 \times 10^{-19}\,\cm^{-2}\,\text{s}^{-1}\,\text{sr}^{-1}$, depending on the fraction of PMBHs that capture protons \cite{Ghosh:1990ki}.  Taking the more conservative limit gives $f_{\sbullet} < 1 \times 10^{3}\,M_{26}$, setting a constraint on masses $\Mewbh < 1 \times 10^{23}$~GeV, a bit stronger than the MACRO constraint.

White dwarf destruction may also set a constraint, potentially adding a secondary constraint to the Parker bound between the annihilation to neutrino constraints and the lensing constraints.  However, a detailed hydrodynamical simulation taking into account the effects of the EWS corona would be required \cite{Montero-Camacho:2019jte}, beyond the scope of this work.

We have not discussed how PMBHs form, instead focusing on the phenomenology if 
they are present.  One example of a formation mechanism is that they may start 
as ordinary primordial BHs, then absorb on average $N$ randomly charged monopoles 
leaving them with a typical charge $\sim \sqrt{N}$, then Hawking radiate until 
they approach near extremality \cite{Stojkovic:2004hz,Bai:2019zcd}.  In such a 
scenario, bounds from the effects of their evaporation on Big Bang 
nucleosynthesis, the cosmic microwave background, and gamma rays could set 
additional constraints \cite{Carr:2009jm}.  Note, these constraints may be 
modified if the BHs obtain a large enough charge to form a corona 
before Hawking radiating to near-extremal.  If, on the other hand, PMBHs are 
born extremal or near extremal, these constraints are relaxed.

If PMBHs are indeed primordial, then they can form binaries in the early Universe that merge today, giving high energy neutrinos and gamma rays throughout the sky.  An estimate of this signal is given in \cite{Bai:2019zcd}, but more detailed numerical work is needed, particularly on binary disruption (see, \eg, \cite{Raidal:2018bbj,Vaskonen:2019jpv}).  Alternatively, binaries may form in galactic halos, but the merger rate from this population of binaries is smaller than the merger rate of primordially-formed binaries \cite{Ali-Haimoud:2017rtz}.

To conclude, PMBHs are interesting long-lived objects that require no new physics beyond the SM and general relativity.  We have outlined many search strategies and shown the PMBH abundance is already relatively constrained compared to dark matter.  Nevertheless, they remain an interesting target for future searches.  In particular, PMBH mergers or baryon number violating processes offer the possibility to detect Hawking radiation.  Furthermore, this Hawking radiation would be emitted as $2d$ modes from the electroweak-symmetric corona, whose spectrum may be differentiated from ordinary $4d$ Hawking radiation.  If a signal is observed, this distinction could provide strong evidence for the PMBH interpretation.

\subsubsection*{Acknowledgements}
The work of YB, MK, and NO is supported by the U.S.~Department of Energy under the contract DE-SC-0017647.  The work of JB is supported by PITT PACC.

\begin{appendix}
\section{Dirac equations and $2d$ modes}
\label{sec:dirac-equations}

In this section, we follow Ref.~\cite{Maldacena:2018gjk} to discuss the solutions to the Dirac equation in a background BH geometry and magnetic field. Rather than only considering the massless case in \cite{Maldacena:2018gjk}, we also keep the fermion mass in our discussion. For a general metric in spherical coordinates, 
\beqa
\label{eq:metric-2}
ds^2 = e^{2\sigma(t, x)}\,\left( - dt^2 + dx^2\right) + R^2(t, x)\, \left( d\theta^2 + \sin^2{\theta}\,d\phi^2\right) ~, 
\eeqa
and $A_\phi = \frac{Q}{2}\,\cos{\theta}$. For extremal BHs, the above metric is related to the one in the ordinary spherical coordinate in \eqref{eq:metric-1} by 
\beqa
\label{eq:x-r-relation}
dx = \frac{dr}{f(r)}  \,, \qquad \, e^{2\sigma(t, x)} = f(r) \equiv \left(1 - R_{\rm e}/r\right)^2 \,, \qquad R(t, x) = r  ~.
\eeqa
Choosing the gamma matrices in the spinor 
representation~\cite{Freedman:2012zz}, 
\beqa
\label{eq:gamma-matrix-spinor}
\widetilde{\gamma}^0 = i\sigma_x \otimes \mathbb{I}_2 \,, \quad \widetilde{\gamma}^1 = \sigma_y \otimes \mathbb{I}_2 \,,\quad \widetilde{\gamma}^2 = \sigma_z \otimes \sigma_x \,,\quad \widetilde{\gamma}^3 = \sigma_z \otimes \sigma_y ~, 
\eeqa
the four-dimensional spinors can be written as a tensor product of two 
dimensional spinors $\widetilde{\chi}_{\alpha \beta} = \psi_\alpha \otimes 
\eta_\beta$.

The Dirac operator in the bi-spinor representation is 
\beqa
\slashed{D} = e^{-\sigma}\Big{[}i\sigma_{x}\left(\partial_{t}+\frac{\dot{\sigma}}{2}\right) + \sigma_{y}\left(\partial_{x}+\frac{\sigma'}{2}+\frac{R'}{R}\right)\Big{]}\otimes \mathbb{I}_2  
+ \frac{\sigma_{z}}{R} \otimes \Big{[}\sigma_{y}\frac{\partial_{\phi}-i A_{\phi}}{\sin\theta} + \sigma_{x}\left(\partial_{\theta}+\frac{\cot\theta}{2}\right)\Big{]} ~.
\eeqa
Here, $\dot{\sigma} = \partial \sigma / \partial t$, $\sigma'= \partial \sigma /\partial x$ and $R' = \partial R / \partial x$. 
Using the ansatz with separation of variables 
\beqa
\label{eq:chi-tilde-solution}
\widetilde{\chi}_{\alpha \beta} = \frac{e^{-\frac{1}{2}\sigma}}{R}\, \psi_\alpha(t, x)\,\eta_\beta(\theta, \phi) ~,
\eeqa
the Dirac equation $\slashed{D}  \widetilde{\chi} = m_\chi \, \widetilde{\chi}$ becomes
\beqa
\label{eq:eta}
\Big{[}\sigma_{y}\frac{\partial_{\phi}-i A_{\phi}}{\sin\theta} + \sigma_{x}\left(\partial_{\theta}+\frac{\cot\theta}{2}\right)\Big{]} \eta &=&  0 \,, \\
\label{eq:psi}
\left( i \sigma_x \partial_t + \sigma_y \partial_x\right) \psi = m_\chi \, e^\sigma \psi ~.  
\eeqa
Eq.~\eqref{eq:eta} can be solved exactly with the solution for $Q > 0$ given as~\cite{Maldacena:2018gjk}
\beqa
\eta_1 &=& 0 ~, \\
\eta_2 &=& \left(\sin{\frac{\theta}{2}}\right)^{j-m}\,  \left(\cos{\frac{\theta}{2}}\right)^{j+m}\,e^{im\phi} = \frac{ (1-\cos{\theta})^{\frac{q - m}{2}} \, (1+\cos{\theta})^{\frac{q + m}{2}}}{2^{q-\frac{1}{2}}\,(\sin{\theta})^{\frac{1}{2}}}\, e^{im\phi} ~,
\eeqa
with $j = (|Q| - 1)/2 \equiv q - 1/2$ and $ -j \leq m \leq j$. For $Q < 0$, one can switch 
$\eta_1 \leftrightarrow \eta_2$. For $Q=0$, there is no solution. If $m_\chi=0$, there are $Q$ two-dimensional massless spinor modes. The forms of $\eta_1$ or $\eta_2$ depend on the gauge choice. If we choose a different gauge with $A_\phi = \frac{Q}{2}(1 - \cos{\theta})$, the solution for $Q >0$ is
\beqa
\label{eq:eta-1-solution}
\eta_1 &=& \, \left(\sin{\frac{\theta}{2}}\right)^{j+m}\,  \left(\cos{\frac{\theta}{2}}\right)^{j-m}\,e^{i(q + m)\phi}= \frac{(1-\cos{\theta})^{\frac{q + m}{2}} \, (1+\cos{\theta})^{\frac{q - m}{2}} }{2^{q-\frac{1}{2}}\,(\sin{\theta})^{\frac{1}{2}}}\,e^{i(q+m)\phi}~,\\
\label{eq:eta-2-solution}
\eta_2 &=& 0 ~.
\eeqa

The solution for $\eta_{1,2}$ is related to the spin-weighted spherical harmonics or the monopole harmonics  ${}_q Y_{lm}$ with $l = q \equiv |Q|/2$~\cite{spin-harmonics-math,Wu:1976ge}, which is given by
\beqa
\label{eq:spin-harmonics}
{}_q Y_{q, m}(\theta, \phi) = M_{q, q, m} \, \frac{(-1)^{q+m} (2 q)!}{2^{q+m}(q+m)! (q-m)!}\,(1-\cos{\theta})^{\frac{(q+m)}{2}} \,(1+\cos{\theta})^{\frac{(q-m)}{2}} \,e^{i(q+ m)\phi} ~, 
\eeqa
where the normalization factor $M_{q, q, m}  = 2^m [(2 q + 1)(q - m)! (q + m)!/(4\pi (2q)!)]^{1/2}$~\cite{Wu:1976ge}. For the gauge choice of $A_\phi = \frac{Q}{2}(1 - \cos{\theta})$, the relation is 
\beqa
&&\eta_1(\theta, \phi) = C_{q, m-1/2}\, \left(\cos{\textstyle{\frac{\theta}{2}}}\right)^{-1} \, e^{i \frac{1}{2} \phi} \, {}_q Y_{q, m-1/2}( \theta, \phi) ~, \\
\mbox{or} &&  \eta_1(\theta, \phi) = C_{q, m+1/2}\, \left(\sin{\textstyle{\frac{\theta}{2}}}\right)^{-1} \, e^{- i \frac{1}{2} \phi} \, {}_q Y_{q, m+1/2}( \theta, \phi) ~. 
\eeqa
Using \eqref{eq:spin-harmonics} for ${}_q Y_{q, m-1/2}$ and ${}_q Y_{q, m+1/2}$, one has the following relation
\beqa
\label{eq:Cq-relation}
\frac{C_{q, m+1/2}}{C_{q, m-1/2}} = - \left( \frac{q + m + \frac{1}{2} }{q - m + \frac{1}{2}} \right)^{1/2} ~.
\eeqa

As a consistency check, we can compare the result in \eqref{eq:chi-tilde-solution} with $\eta_{1,2}$ given by \eqref{eq:eta-1-solution} and \eqref{eq:eta-2-solution} with the results of 
Ref.~\cite{Kazama:1976fm},  which discusses a similar problem neglecting the 
curvature of spacetime due to the monopole. The solution using the gamma matrices in the spinor basis via \eqref{eq:gamma-matrix-spinor} is  
\beqa
\label{eq:solution-Maldacena}
\widetilde{\chi}  = \left\{d_1\,\frac{e^{i E r}}{r},~ 0 ,~ d_2\,\frac{e^{- i E r}}{r},~ 0   \right\}^T \, \times\,\eta_1 ~,
\eeqa
with $d_{1,2}$ as normalization factor. The solution in Ref.~\cite{Kazama:1976fm} is calculated based on the Dirac basis for the gamma matrix with 
\beqa
\label{eq:gamma-matrix-Dirac}
\gamma^0 =  i\sigma_z \otimes \mathbb{I}_2  \,, \quad \gamma^1 = - \sigma_y 
\otimes \sigma_x \,,\quad \gamma^2 = -\sigma_y \otimes \sigma_y \,,\quad 
\gamma^3 = -\sigma_y \otimes \sigma_z  ~, 
\eeqa
where we have multiplied an additional factor of $i$ for the Dirac matrices in Ref.~\cite{Kazama:1976fm} to match the metric convention in Ref.~\cite{Maldacena:2018gjk}. For the ``type (3)" solution in Ref.~\cite{Kazama:1976fm}, one has 
\beqa
\label{eq:solution-Kazama}
\chi = \begin{pmatrix}
	- f(r)\, \left(\frac{j-m+1}{2 \, j +2}\right)^{1/2} \, {}_{q}Y_{q,m-1/2} \\
	f(r) \left(\frac{j+m+1}{2 \, j +2}\right)^{1/2} \, {}_{q}Y_{q,m+1/2} \\
	- g(r)\, \left(\frac{j-m+1}{2 \, j +2}\right)^{1/2} \, {}_{q}Y_{q,m-1/2} \\
	g(r) \left(\frac{j+m+1}{2 \, j +2}\right)^{1/2} \, {}_{q}Y_{q,m+1/2} 
	\end{pmatrix} ~,
\eeqa
with $f(r) = (2/\pi)^{1/2}\,\sin{(E r + \delta_3)}/(E r)$ and $g(r) = -i\,(2/\pi)^{1/2}\,\cos{(E r + \delta_3)}/(E r)$. Here, $j = q - 1/2$ with $q> 0$ and $m = -j, \cdots, j$. 

To match the two solutions in \eqref{eq:solution-Maldacena} and \eqref{eq:solution-Kazama}, there are two relevant unitary transformations. The first one is the transformation for the different choices of Dirac matrices.  The two bases in \eqref{eq:gamma-matrix-spinor} and \eqref{eq:gamma-matrix-Dirac} are related to each other by a unitary transformation, $\widetilde{\gamma}^\mu = U\, \gamma^\mu \, U^\dagger$, with  
\beqa
	U = \frac{1}{2}\, 
	\begin{pmatrix}
		1 & 1& 1 & 1  \\
		1 & -1& -1 & 1  \\
		1 & 1& -1 & -1  \\
		1 & -1 & 1 & -1  
	\end{pmatrix} ~. 
\eeqa

The second unitary matrix is a rotation of the tangent space relative to 
	the coordinate space. In \cite{Maldacena:2018gjk}, the tangent space is chosen in the convention such 
	that the $x$-axis aligns with the coordinate radial direction. This rotation 
	acts on 
	spinors as
	\begin{equation}
	R_{1/2} = \frac{1}{\sqrt{2}} e^{i \, \pi/4} \, 
	\begin{pmatrix}
	e_- \, c + e_+ \, s & 0 & 0 & 
	i \,(e_+ \, c - e_- \, s) \\
	0 & -i \,(e_+ \, c + e_- \, s) & e_- \, c - e_+ \, s & 0 \\
	0 & i \,(e_+ \, c - e_- \, s) & 
	e_- \, c + e_+ \, s & 0 \\
	e_- \, c - e_+ \, s & 0 & 0 & 
	-i \,(e_+ \, c + e_- \, s)
	\end{pmatrix}\, ,
	\end{equation}
	where $c = \cos{\frac{\theta}{2}}$, $s = \sin{\frac{\theta}{2}}$, and $e_\pm = e^{\pm \, i \, 
		\phi/2}$.  The two solutions in \eqref{eq:solution-Maldacena} and \eqref{eq:solution-Kazama} are related by
\beqa
\label{eq:relation-chi}
\widetilde{\chi} = R^\dagger_{1/2}\, U\, \chi ~. 
\eeqa
Using the relation in \eqref{eq:Cq-relation}, the above equation becomes
\beqa
 \left\{d_1\,\frac{e^{i E r}}{r},~ 0 ,~ d_2\,\frac{e^{- i E r}}{r},~ 0   \right\}^T \, \times\,\eta_1   &=& \nonumber \\
 &&\,\hspace{-5cm}  -\,e^{-i \frac{\pi}{4} } \, \sqrt{\frac{1}{2} - \frac{m}{1+ 2q} } \, C_{q, m-1/2}^{-1}\,  \left\{\frac{f(r)+g(r)}{\sqrt{2}}, ~ 0 ,~ \frac{f(r) - g(r)}{\sqrt{2}},~ 0   \right\}^T \, \times\,\eta_1 ~. 
\eeqa
One can explicitly see that the $\theta$- and  $\phi$-dependences  on the both sides of the equation come from the same function $\eta_1$ and the $r$-dependence is also identical. 

After demonstrating the consistency of solutions in the literature, one can ask which complimentary states must be added to the $2d$ modes to form a complete basis. One can 
use the other two types of solutions in \eqref{eq:solution-Kazama} and perform 
a transformation via \eqref{eq:relation-chi} to obtain the correponding 
solutions in the basis choice of \eqref{eq:solution-Maldacena}. Doing so, one 
could derive the equation of motion for the radial functions by including the 
curved spacetime and obtain the solutions for a Dirac fermion in the MeBH 
background. 

To develop a qualitative understanding of the $2d$ neutrino modes, note that the neutrino has a non-zero hypercharge inside the EWS corona and zero electric charge outside. One could model this situation by choosing a step-function-like $A_\phi = \frac{Q}{2}\cos{\theta} \, \Theta(R_{\rm EW} - r)$ (assuming the boundary condition is spherical, which is only approximately true for a large $Q$). One can explicitely check that a linear combination of solutions for $Q>0$ and $Q<0$ $2d$ modes can have a standing wave solution inside $R_{\rm EW}$ with vanishing wave-function outside. This type of ``particle-hole state" has been discussed for the bosonization~\cite{Coleman:1974bu}, and we leave the more detailed calculation to future exploration. 

Finally, we also comment on the situation for the massive fermion case. For a non-zero mass $m_\chi \neq 0$, Eq.~\eqref{eq:psi} cannot be solved analytically. Defining 
\beqa
\psi = e^{-i E t} 
\begin{pmatrix}
	\psi_1(x) \\
	\psi_2(x)
\end{pmatrix} ~,
\eeqa
Eq.~\eqref{eq:psi} becomes
\beqa
e^{-\sigma}( E - i \partial_x) \psi_2 = m_\chi\, \psi_1  \,, \qquad 
e^{-\sigma}( E + i \partial_x) \psi_1 = m_\chi \, \psi_2   \,,
\eeqa
which can be combined and become
\beqa
[e^{-\sigma}( E - i \partial_x)] [e^{-\sigma}( E + i \partial_x)] \psi_1 = m_\chi^2 \psi_1~. 
\eeqa
Using the relation between $x$ and $r$ in \eqref{eq:x-r-relation}, we convert the above equation into
\beqa
- \left(\partial_r^2 \, + \frac{f'}{2\,f}\,\partial_r\right)\psi_1(r) = \frac{1}{f^2}\left( E^2 + \frac{1}{2}i E\,f' - f\,m_\chi^2 \right) \,\psi_1(r) ~,
\eeqa
with $f(r) = (1 - R_{\rm e}/r)^2$ and $f'(r)\equiv df(r)/dr$. Note that when $r \rightarrow \infty$, the equation provides a simple plane wave solution, $\psi_1(r) = e^{i k r}$, with the normal dispersion relation $k^2 = E^2  - m_\chi^2$ for $E \geq m_\chi$. Close to the event horizon or $r \rightarrow R_{\rm e}$, the mass term is less important. So, effectively, one could treat the system as a particle with a location-dependent mass, or equivalently with some ``attractive potential". For $E < m$, bound state solutions are anticipated. For instance, when a nucleon scatters off the EWS corona region, its mass is reduced and the bound-state-mediated scattering can increase the scattering cross section to the geometric one~\cite{Ponton:2019hux}.

\section{Stopping of a finite-sized PMBH by a plasma}
\label{sec:stopping}

We modify the treatment of monopole stopping in a plasma in \cite{1985ApJ...290...21M}  (see also \cite{Hamilton:1984rh}) to account for a finite-sized monopole.  Since the properties of the plasma inside the 
EW radius $R_{\rm EW}$ are uncertain, we conservatively assume 
that modes at impact parameter less than $R_{\rm EW}$ do not contribute to stopping.  To do so, we approximate the PMBH as a charged spherical 
shell $R$.  Any interactions with impact parameter smaller than this are neglected.  It is simple to verify that the stopping power by particles with impact parameter smaller than $R$ that are absorbed by the BH is insufficient to stop the PMBH.  

Note that the treatment in \cite{1985ApJ...290...21M} neglects non-linearities in the plasma response, however non-linearities are important for the Sun.  Thus, this calculation does not give an accurate result for the stopping power in the Sun.  Still, it gives the correct proportionality of the stopping power on the parameters in the problem.  As we will see, the stopping power monotonically increases with $Q$ even when finite-size effects are included, and the minimum $Q$ for which PMBHs are stopped in astrophysical bodies like the Sun is well within the point-particle regime.  This leads us to conclude that finite-size effects do not affect whether or not PMBHs are stopped.

The charge and current densities for a PMBH with radius $R$, velocity $\mathbf{V}$, and magnetic charge $h_Q$ can be taken as, 
\begin{equation}
\rho(\mathbf{r},t) = \frac{h_Q}{4\pi R^2} \delta(|\mathbf{r}-\mathbf{V}t|-R)\, , \; \; \; \mathbf{J}(\mathbf{r},t)=\mathbf{V} \rho(\mathbf{r},t)  \, .
\end{equation}
We take the Fourier transform of the magnetic charge density to obtain
	\begin{equation}
	\rho(\mathbf{k},\omega) = 2 \, \pi \, h_Q \, \delta(\mathbf{k}\cdot 
	\mathbf{V} - \omega) \, j_0(k \, R) ~,
	\end{equation}  
	where $j_0(x) = \sin x / x$ is a spherical Bessel function, which
	goes to $1$ for $x \ll 1$ and falls off as $1/x$, up to 
	oscillation, at large $x$.
		
The magnetic field in Fourier space is
\begin{equation}
\mathbf{B}(\mathbf{k},\omega) = \mu_0\,i \,\frac{\omega\, \epsilon_{\rm T}\, \mathbf{J} - \mathbf{k} \,\rho}{k^2 - \epsilon_{\rm T} \, \omega^2} \, ,
\end{equation}
where all quantities on the right are assumed Fourier transformed.  The transverse plasma dispersion $\epsilon_{\rm T}$ is given in \cite{1985ApJ...290...21M}.  In the limit $z=\omega/(k\, v_\text{th}) = \mathbf{k}\cdot\mathbf{V} /(k\,v_\text{th}) \ll 1$ with $v_\text{th}$ the thermal velocity of the plasma particles,
\begin{equation}
\epsilon_{\rm T} (k, \omega) \sim 1 + i\,\pi^{1/2}\,(\omega_p^2/\omega^2)\, z \, ,
\end{equation}
with the squared plasma frequency $\omega_p^2=4\pi n_e e^2/m_e$ for electrons with number density $n_e$ and mass $m_e$.  Then, the PMBH power loss is
\begin{align}
\frac{dW}{dt} &= \int d^3 r \, \mathbf{J}(\mathbf{r},t) \cdot \mathbf{B}(\mathbf{r},t)
\\
&= -\frac{\mu_0 \, i \,V\, h_Q^2}{(2 \pi)^3} \int d^3 k \, \left. k_z \, \frac{1-V^2 \, 
\epsilon_{\rm T}}{k^2 - V^2\, \epsilon_{\rm T}\, k_z^2} \,[j_0 (k R)]^2 \right|_{\omega = k_z V}
\\
&= -\frac{\mu_0\, V\, h_Q^2}{4 \pi^2 l^2} \int_0^\infty dk \, k^3\, [j_0(k R)]^2 
\int_{-1}^{1} du \, \frac{1-u^2}{k^4+u^2/l^4} \, ,
\label{eq:dWdt-integral}
\end{align}
where the last line is in the limit $V \ll 1$.
The characteristic attenuation length is
\begin{equation}
l = \pi^{-1/4} (v_\text{th}/V)^{1/2} \omega_p^{-1} \approx 3 \times 10^{-6}~\text{cm} \, .
\end{equation}
The last equality gives the value for the Sun with $V=10^{-3}$. The integral is truncated at finite $k_\text{max}$ to avoid a logarithmic divergence when $R=0$.  This corresponds to the assumption of a linear plasma response in $\epsilon_{\rm T}$, which neglects short-range interactions.  It is taken as related to the distance where the electrostatic and thermal energies are equal: $k_\text{max} \sim 4 \pi\,n\, L_{\rm D}^2 = T^2 / e^2 \approx (2 \times 10^{-9}~\text{cm})^{-1}$, with $n$ the number density of particles in the plasma and $L_{\rm D}$ the Debye length.  However, the result is only logarithmically dependent on this.  Then, the $R=0$ result is
\begin{equation}
\label{eq:stopping-power-pointlike}
\left. \frac{dW}{dt} \right|_{R=0} = - \frac{\mu_0 V h_Q^2}{3 \pi^2 l^2} \left[ \log (k_\text{max}\,l) + \frac{2}{3} \right] \, .
\end{equation}

To generalize to finite radius, we must now account for the $j_0(kR)$ term. We approximate the integral in (\ref{eq:dWdt-integral}) by taking $j_0(kR)=1$ for $k \in [0,1/R)$ and $j_0(kR)\sim 1/(kR)$ for $k \in [1/R,\infty)$ (ignoring the sine dependence).  For $R \ll l$, the result is the same as \eqref{eq:stopping-power-pointlike} but with $k_\text{max}$ replaced by $k_\text{max}'=\min(k_\text{max},1/R)$, while for $R \gg l$ and $k_\text{max} \gg 1/R$
\begin{equation}
\frac{dW}{dt} = -\frac{\mu_0 V h_Q^2}{4 \pi R^2} \left[\log\left(\frac{R}{l }\right)+\frac{1}{4}\right]
\, .
\end{equation}

The radius  is  the EWS corona radius, 
\begin{align}
R_\text{EW} &= \sqrt{\frac{Q}{2}} \frac{1}{m_h} \approx (10^{-8}~\text{cm}) \sqrt{\frac{Q}{10^{16}}} ~ .
\end{align}
Thus, the stopping length $L_{\rm S} \sim \frac{1}{2} \Mewbh V^2 (dW/dx)^{-1} \propto Q V^3 (dW/dt)^{-1}$ is monotonic with $Q$, since 
\begin{equation}
\frac{dW}{dt} \propto \left\{ 
\begin{array}{ll}
 Q^2 \, , \; \; \; & k_\text{max} R \ll 1 ~,
\\
 Q^2 \, \log (1/\sqrt{Q}) \, , \; \; \; & k_\text{max} R \gg 1 ~\&~ R=R_\text{EW} \ll  l ~,
\\
 Q \log (\sqrt{Q}) \, , \; \; \; & k_\text{max} R \gg 1~\&~R=R_\text{EW} \gg 
l~(\textit{i.e.,}~Q \gg 10^{16}~\&~Q \lesssim Q_\text{max}) ~.
\end{array}
\right. 
\nonumber
\end{equation}
Note $L_{\rm S}$ is monotonically decreasing as $Q$ increases regardless of how $k_\text{max}$, $R$, and $\ell$ are related.

Finally, note that in the point-like MeBH case, the stopping length is smaller than the solar radius for $Q \gtrsim \Qstopsun$ (see below).  For this charge, the point-like approximation is valid.  Since the stopping length monotonically decreases with $Q$ even after accounting for finite size effects, all extremal magnetic BHs above this charge are stopped.

\subsection{Stopping in other materials}
\label{sec:stop-other}

To a rough approximation, the stopping power in other materials like conductors, insulators, or degenerate gases is very similar the stopping power of plasmas considered in the previous section, but with the thermal velocity of electrons $v_\text{th}$ replaced by the Fermi velocity $v_{\rm F}$ \cite{Ahlen:1982mx}.  In other words, from the point-like approximation in (\ref{eq:stopping-power-pointlike}),
\begin{equation}
\frac{dW}{dx} \sim \frac{\omega_p^2 \,h_Q^2 \,V}{v_{\rm F}} \sim \frac{n_e \, e^2 \, h_Q^2 \,V}{v_{\rm F} \,m_e} \, .
\end{equation}
This can be used for objects like neutron stars and the Earth. The minimum 
charge for a PMBH to be captured while traveling through a body is then 
estimated by requiring
	\begin{equation} 
		\frac{\Mewbh \, v^2/2}{ dW/dx|_{V = (v^2 + v_{\rm esc}^2)^{1/2}}} 
		\gtrsim R ~ , 
	\end{equation}
	where $v \approx 10^{-3}$ is the velocity of the PMBH far from the body, $v_\text{esc}$ is the body's escape velocity, and 
	$R$ is the radius of the body.  If this condition is satisfied, then the 
	PMBH becomes gravitationally bound to the body and will, either during the first crossing or over the course of 
	further crossings, stop inside the body. The resulting minimal charge is
	\begin{equation}
	\label{eq:Qstopmin}
	Q_{\rm stop,min} \sim \frac{c_W \, v^2 \, v_{\rm F} \, m_e \, M_{\rm pl}}{8 \, \pi^{3/2} \,e \, 
	n_e \, R \, (v^2 + v_{\rm esc}^2)^{1/2}} ~,
	\end{equation}
which assumes that once the PMBH is gravitationally bounded, it will be quickly captured, usually during its first pass through the object (this is true for the Sun, the Earth and neutron stars, which provide non-trivial constraints). 
We provide relevant quantities in Table \ref{tab:quantities}.  For stars, the proton-to-nucleon ratio (equivalent to the electron-to-nucleon ratio) is $Z/A \sim 1$, while for the Earth $Z/A \sim 1/2$.

For the Sun, we use the more precise calculation of the stopping power in \cite{Frieman:1985dv} (though with a factor of two larger stopping power \cite{Ahlen:1996ax}), which gives 
$Q_\text{stop,min} \sim \Qstopsun$, compared to $Q_\text{stop,min} \sim 390$ using (\ref{eq:Qstopmin}) and the quantities in Table \ref{tab:quantities}.

\begin{table}[h!]
\renewcommand{\arraystretch}{1.5}
	\label{tab:quantities}
	\caption{Physical quantities relevant for stopping. }
	\begin{tabular}{c|c|c|c}
	\hline\hline
		& $n_e$ & electron $v_\text{th}$ or $v_{\rm F}$ & $Q_\text{stop,min}$
		\\ \hline
		Sun & 10$^{24}$ cm$^{-3}$ & $v_\text{th}=0.058$  (from $T=10^7$ K) & \Qstopsun~\cite{Frieman:1985dv,Ahlen:1996ax}
		\\ \hline
		Earth & $(5.5~\text{cm}^{-3})\,\frac{Z}{A}\,N_A \sim 1.7\times 10^{24}~\text{cm}^{-3}$ & $v_{\rm F} \sim \sqrt{\frac{2 (1~\text{eV})}{(0.511~\text{MeV})} } \sim 2 \times 10^{-3}$ & \Qstopearth
		\\ \hline
		Neutron star & $6 \times 10^{37}~\text{cm}^{-3}$ & $v_{\rm F} \sim 1$ & 1
		\\ \hline
		White dwarf & $6 \times 10^{29}~\text{cm}^{-3}$ & $v_{\rm F} \sim 0.7$ & 1 \\
  \hline 
 \hline
	\end{tabular}
\end{table}

\end{appendix}
\setlength{\bibsep}{6pt}
\bibliographystyle{JHEP}
\bibliography{PMBH}
\end{document}